\documentclass[%
 reprint,
 amsmath,amssymb,
 aps,
superscriptaddress,
]{revtex4-2}

\usepackage{graphicx}
\usepackage{dcolumn}
\usepackage{bm}
\usepackage{multirow}
\usepackage{array}
\usepackage{bm}
\usepackage{color}

\begin{document}

\title{N\'eel vector-dependent anomalous transport in altermagnetic metal CrSb} 

\author{Tianye Yu}
\thanks{These authors contributed equally to this work.}
\affiliation{Shenyang National Laboratory for Materials Science, Institute of Metal Research,Chinese Academy of Sciences, Shenyang 110016, China.}

\author{Ijaz Shahid}
\thanks{These authors contributed equally to this work.}
\affiliation{Shenyang National Laboratory for Materials Science, Institute of Metal Research,Chinese Academy of Sciences, Shenyang 110016, China.}
\affiliation{School of Materials Science and Engineering, University of Science and Technology of China, Shenyang 110016, China.}
\author{Peitao Liu}
\affiliation{Shenyang National Laboratory for Materials Science, Institute of Metal Research,Chinese Academy of Sciences, Shenyang 110016, China.}
\affiliation{School of Materials Science and Engineering, University of Science and Technology of China, Shenyang 110016, China.}

\author {Ding-Fu Shao}
\email{dfshao@issp.ac.cn}
\affiliation {Key Laboratory of Materials Physics, Institute of Solid State Physics, Hefei Institutes of Physical Science, Chinese Academy of Sciences, Hefei 230031, China}




\author{Xing-Qiu Chen}
\email{xingqiu.chen@imr.ac.cn}
\affiliation{Shenyang National Laboratory for Materials Science, Institute of Metal Research,Chinese Academy of Sciences, Shenyang 110016, China.}
\affiliation{School of Materials Science and Engineering, University of Science and Technology of China, Shenyang 110016, China.}

\author{Yan Sun}
\email{sunyan@imr.ac.cn}
\affiliation{Shenyang National Laboratory for Materials Science, Institute of Metal Research,Chinese Academy of Sciences, Shenyang 110016, China.}
\affiliation{School of Materials Science and Engineering, University of Science and Technology of China, Shenyang 110016, China.}

\begin{abstract}
Altermagnets are predicted to exhibit 
anomalous transport phenomena, such as the anomalous Hall 
and Nernst effects, as observed in ferromagnets but with a 
vanishing net magnetic moment, akin to antiferromagnets. 
Despite their potential, progress has been limited due to the scarcity of metallic altermagnets. Motivated by the recent discovery of the altermagnetic metal CrSb, we conducted a systematic study of its electrical and thermoelectric transport properties, using first-principles calculations. CrSb exhibits low magnetocrystalline anisotropy energy, enabling the manipulation of the N\'eel vector in CrSb films through a suitable ferromagnetic substrate. The anomalous Hall and Nernst conductivities reach their maximum when the Néel vector is aligned along $\frac{1}{2}$\textbf{\textit{a}}+\textbf{\textit{b}}. The 
origins of both conductivities were analyzed in terms of 
Berry curvature distribution. Our results demonstrate that CrSb provides a good platform for investigating the N\'eel vector-dependent anomalous transport in altermagnetic metals.

\end{abstract}

\pacs{Valid PACS appear here}
\maketitle
\section{INTRODUCTION}
Magnetism has long been a central topic in condensed-matter physics. Within the domain of collinear magnetism, attention has traditionally been directed toward two primary types: ferromagnetism, where spins in a lattice align in a uniform direction, and antiferromagnetism, characterized by two spin sublattices with opposite orientations. Recently, a novel class of magnetism, termed altermagnetism, has been proposed\cite{altermagnetism1,altermagnetism2,altermagnetism3,altermagnetism4,altermagnetism5,altermagnetism6,altermagnetism7}. This emerging class distinguishes itself from the two conventional types.
In altermagnets, the net magnetization is vanishing, similar 
to antiferromagnets. However, it is the rotational symmetry
(proper or improper, symmorphic or nonsymmorphic) not the 
inversion or translation symmetry operation  that relates 
the two antiparallel sublattices, leading to a spin-splitting 
band structure, reminiscent of ferromagnets. As a key 
characteristic of altermagnets, the spin-splitting electronic 
structure has been widely investigated through theoretical 
calculations\cite{science_advance_RuO2, liu2021,arxive-correlated-altermagnets,crystal-thermal-transport,DFT-CaCrO3-AHC,DFT-strain-ReO2,high_througput_spin_splitting_DFT_calculations,magnon-RuO2-PRL,minimal-models,moke_RuO2,Orbital-selective-and-correlation-altermagnetism,PRB-Tables2,PRM-tables,space_group_62,topological-superconductivity1,topological-superconductivity2,topological-transition,zhanjie_RuO2,zhongyi_LU_Nb2FeB2,Zhongyi-Lu-AI,zhongyi-Lu-altermagnetic-FE,Zhongyi-Lu-NiF3,liu2021,liu2024,FeSb2}, 
but experimental evidence remains scarce. The first 
experimentally confirmed altermagnet, identified through 
angle-resolved photoemission spectroscopy (ARPES) measurements, 
is MnTe\cite{MnTe-ARPES1-Nature,MnTe-ARPES2-PRL,MnTe-ARPES3-PRb}.

Owing to the odd parity of Berry curvature (BC) with respect to 
time-reversal operation, nonmagnetic systems with time-reversal symmetry exhibit no anomalous Hall effect (AHE), whereas the situation is different for 
ferromagnets\cite{AHC2,Xiao-Di-RMP}. 
In certain magnetic materials, while time-reversal 
symmetry is broken, the combination of time-reversal 
symmetry with specific space group operation ensures that 
the net BC over the entire Brillouin zone sums to zero\cite{BC1,BC2}. This  generally 
applies to antiferromagnets with two magnetic sublattices
connected by $PT$ or ${T|t}$ symmetry, where $P$ is inversion, $T$ is time reversal, and $t$ is a lattice translation. However, a nonzero net BC and a finite intrinsic AHE are permitted in altermagnets with spin splitting. The AHE in altermagnets has been both theoretically 
studied and experimentally observed in RuO$_2$\cite{AHC_EXP_RuO2}, 
Mn$_5$Si$_3$\cite{nc_Mn5Si3}, and MnTe\cite{AHC_EXP_MnTe}. 
However, specific constraints within these materials limit 
further exploration of AHE. To be specific, the magnetic ground 
state of RuO$_2$ remains under debate\cite{RuO2-AM1,RuO2-AM2,RuO2-AM3,RuO2-huang,RuO2-nonmagnetic1,RuO2-nonmagnetic2,RuO2-nonmagnetic3}. 
MnTe is a semiconductor with a band gap of about 
1.3 eV, exhibiting suboptimal conductivity compared to 
metals\cite{MnTe-band_gap1,MnTe-band_gap2}. As for Mn$_5$Si$_3$, 
its magnetic states are sensitive to temperature and noncollinear magnetic structures are involved, which limits the measurements and applications in anomalous 
transport of collinear altermagnetic phase\cite{Mn5Si3-magnetic_state1,Mn5Si3-magnetic_state2,Mn5Si3-magnetic_state3,Mn5Si3-magnetic_state4}.
Therefore, altermagnetic metals with high N\'eel 
temperatures are widely sought after.

Recently, CrSb was identified as a new altermagnetic metal through experimental studies, unveiling a range of novel physical phenomena. The existence of 
different carriers with high mobility yielding a 
nonlinear magnetic field dependence of the Hall 
effect was revealed\cite{CrSb-multicarrier}. 
Additionally, ARPES measurements and first-principles 
calculations identified surface Fermi arcs near the 
Fermi level, which originate from bulk band 
topology\cite{CrSb-fermi-arc}. Most notably, 
the key characteristic of altermagnets, spin 
splitting up to ~1.0 eV along non-high-symmetry 
paths near the Fermi level, was observed by ARPES 
measurements from different  
groups\cite{CrSb-ARPES1,CrSb-fermi-arc,CrSb-ARPES2,CrSb-ARPES3,CrSb-ARPES4}. 
These properties establish CrSb as a prominent material among altermagnets. In comparison to MnTe, CrSb exhibits metallic characteristics that offer superior conductivity, thereby benefiting transport property measurements. Furthermore, in contrast to Mn$_5$Si$_3$, the altermagnetic ground state of CrSb exhibits little sensitivity to temperature, with a relatively high Néel temperature exceeding 700 K\cite{CrSb-Neel-temperature}, thereby offering a broad temperature range for investigating properties related to the altermagnetic state.

Considering the significance of altermagnetic metals in the study of anomalous transport and the remarkable characteristics of CrSb, including its high $T_N$ and large spin splitting, exploring anomalous transport in CrSb is of great importance. In this study, we conducted a comprehensive investigation of the anomalous transport properties in CrSb. First, the magnetocrystalline anisotropy energy (MAE) of CrSb was evaluated based on total energy calculations. Next, we examined the dependence of the anomalous Hall conductivity (AHC) and anomalous Nernst conductivity (ANC) in CrSb on the orientation of the N\'eel vector. Finally, the origins of AHC and ANC were explored by analyzing the BC distribution across the entire Brillouin zone.

\section{METHODS}
We performed density functional theory (DFT) calculations using the Vienna \textit{ab initio} simulation package (VASP)\cite{vasp}. The experimental lattice constants 
a = b = 4.103 $\mathrm{\AA}$ and c = 5.463 $\mathrm{\AA}$ were used, 
which theoretically reproduced the ARPES results very 
well\cite{CrSb-fermi-arc,CrSb-ARPES4}. The exchange-correlation 
functional was treated using the Perdew-Burke-Ernzerhof (PBE)
generalized-gradient approximation\cite{pbe}. An energy cutoff 
of 400 eV was employed, with an energy criterion of $10^{-7}$ eV, 
and a $k$ mesh of 11 × 11 × 9 points was used. From the DFT 
band structure, Wannier functions were generated using 
WANNIER90\cite{wannier90}, with initial projections onto 
the 3$d$ orbitals of Cr and the 5$p$ orbitals of Sb. To evaluate 
the BC, the tight-binding Hamiltonian $H$ was constructed from 
the Wannier functions and applied with the Kubo formula
\begin{equation}
\Omega_{i j}^n=Im\sum_{m \neq n} \frac{\langle n| \frac{\partial H}{\partial k_i}|m\rangle\langle m| \frac{\partial H}{\partial k_j}|n\rangle-(i \leftrightarrow j)}{\left(E_n-E_m\right)^2},    
\end{equation}
where $\Omega_{ij}^n$ denotes the $ij$ component of the BC of 
the $n$th band, $|n \rangle$ and $|m \rangle$ represent the 
eigenstates of $H$, respectively. Based on this equation, 
we can obtain the $ij$ component of the AHC $\sigma_{ij}$ 
by integrating the BC over the entire Brillouin zone
\begin{equation}
\sigma_{i j}=\frac{e^2}{\hbar} \sum_n^{\text {occ }} \int \frac{d^3 k}{(2 \pi)^3} \Omega_{i j}^n,
\end{equation}
where $occ$ denotes the bands occupied by electrons. 
The ANC $\alpha_{ij}$ can be obtained by 
\begin{equation}
\begin{aligned}
\alpha_{i j}= & \frac{1}{T} \frac{e}{\hbar} \sum_n \int \frac{d^3 k}{(2 \pi)^3} \Omega_{i j}^n\left[\left(E_n-E_F\right) f_n\right. \\
& \left.+k_B T \ln \left(1+\exp \frac{E_n-E_F}{-k_B T}\right)\right],
\end{aligned}
\end{equation}
where $T$ denotes the temperature, $k_B$ is the Boltzmann 
constant, $E_F$ is the Fermi level, and $f_n$ is the Fermi 
distribution. A $k$ mesh of 192 $\times$ 192 $\times$ 192 
points was used for both AHC and ANC calculations to obtain 
convergent values.

\section{RESULTS}
\subsection{Spin-splitting band structure of CrSb}
Figure 1(a) presents the three-dimensional crystal 
structure of CrSb, which has a typical hexagonal 
NiAs-type configutation. Within one unit cell, there are 
two Cr atoms  and two Sb atoms. The space group of CrSb 
is $P6_3/mmc$, where each Cr atom is surrounded by six 
Sb atoms. Along the $\textbf{\textit{c}}$ direction, the layers of Cr and 
Sb atoms alternate in stacking. The Cr atoms within the 
same layer have the same spin orientation along the $\textbf{\textit{c}}$ 
direction, while Cr atoms in different layers exhibit 
opposite spin orientations. Considering only the 
two opposite magnetic sublattices, CrSb displays typical 
A-type antiferromagnetism. However, when taking the local environment formed by the Sb atoms into account, 
CrSb behaves as an altermagnet due to the $M_z$ mirror 
operation, or the combination of the rotation operation 
$C_{6z}$ and translation operations relating two sublattices, rather than inversion or translation operation typical in antiferromagnets. In the spin-group formalism, CrSb is a g-wave altermagnet with four nodal planes protected by [$C_2 || C_{6z}t$] and [$C_2 || M_z$], and belongs to $2^6/^1m^2m^1m$ spin Laue group\cite{altermagnetism2}.

The fundamental hallmark of altermagnetism lies in the spin-splitting 
band structure. Figure 1(c) presents the band structure 
along the high-symmetry lines, as well as along a 
non-high-symmetry path, labeled $A-B$, as shown in 
Fig. 1(b). The bands along $\Gamma-M-K-\Gamma$ on 
the $k_z = 0$ plane and $A-L-H-A$ on the $k_z = \pi$ 
plane are spin-degenerate. In contrast, this is not the 
case along the $L-\Gamma$ and $A-B$ path, where point $B$ locates at 
midpoint of $\Gamma-M$, indicating the spin-splitting 
characteristic in CrSb. The calculated band structure agrees with the results reported in Ref \cite{altermagnetism2}, which was the first to identify the altermagnetic band structure of CrSb. Figure 1(d) shows the 
density of states (DOS), where the total and 
orbital-resolved DOS for opposite spins are identical 
in magnitude. This equality arises because the symmetry 
operations connecting the two sublattices also relate 
the energy bands with opposite spins in reciprocal space.

\subsection{Magnetocrystalline anisotropy energy in CrSb}
Despite the spin-splitting characteristics in CrSb, 
the AHE is symmetry-forbidden when the N\'eel vector 
aligned along \textbf{\textit{c}}. Interestingly, the 
calculated MAE is found to 
be rather weak, indicating that the N\'eel vector 
can be easily tuned by an external field. Figure 2 illustrates the dependence of total energy on the Néel vector, with the total energy of the nonmagnetic state normalized to zero. The inset depicts the vector orientations, where \textbf{\textit{c}} indicates the out-of-plane direction. All ferromagnetic 
and altermagnetic states have negative energies, 
and the altermagnetic states are significantly 
lower than the ferromagnetic ones. This result is robust, remaining consistent even when the Hubbard $U$ is applied to the Cr 3$d$  orbitals, and it aligns well with experimental findings that confirm the altermagnetic state as the magnetic ground state\cite{CrSb-Neel-temperature}. Moreover, the difference of total energies among different 
ferromagnetic or altermagnetic states with different 
magnetic moments and N\'eel vector orientations are 
too small to be discernible in Fig. 2. Table I lists 
the total energies of altermagnetic states calculated 
with different N\'eel vector. The difference of total 
energies are smaller than 1 meV, approaching the accuracy limit of DFT. 
Besides, except for the case with $U$ = 0.75 eV, the altermagnetic state 
with N\'eel vector aligned along $\textbf{\textit{c}}$ 
turns to be the magnetic ground state, the same as 
experimental findings\cite{CrSb-Neel-temperature}. Additionally, calculations with different N\'eel vector orientations, without using a penalty functional to constrain the magnetic moment, easily converge to the given orientation of the N\'eel vector, further confirming the weak MAE of CrSb. Within this context, it is meaningful to study the dependence of anomalous 
transport on the orientation of the N\'eel vector.   
\subsection{Anomalous Hall and Nernst conductivities}
Figure 3(a) presents the nonzero elements of the AHC 
tensor as the N\'eel vector rotates from  $\textbf{\textit{c}}$ to three representative orientations, 
within the corresponding plane formed by the N\'eel vector present in the legend and \textbf{\textit{c}}. 
Notably, using the Cartesian coordinates defined in Fig. 1(a), there are two 
independent nonzero elements of the AHC tensor, namely 
$\sigma_{xz}$ and $\sigma_{yz}$ when the N\'eel vector 
rotates from \textbf{\textit{c}} to \textbf{\textit{a}}+\textbf{\textit{b}}, and 
$\sigma_{xy}$ and $\sigma_{xz}$ when it rotates from 
\textbf{\textit{c}} to $\frac{1}{2}$\textbf{\textit{a}}+\textbf{\textit{b}}. This is 
consistent with symmetry analysis. For example, When the Néel vector is oriented between \textbf{\textit{c}} and $\frac{1}{2}$\textbf{\textit{a}}+\textbf{\textit{b}}, the presence of the combined symmetry of \{$m_z$ $\mid$ 0 0 $\frac{1}{2}$\} and the combination of \{C$_{2y}$ $\mid$ 0 0 $\frac{1}{2}$\} with the time-reversal operation $T$ implies that $\sigma_{yz}\,=\,0$. Specifically, when considering the C$_{2y}$ symmetry operation, the three components of the Berry curvature [$\Omega_{yz}$($k_x, k_y, k_z$), $\Omega_{zx}$($k_x, k_y, k_z$), $\Omega_{xy}$($k_x, k_y, k_z$)] transform to [$-\Omega_{yz}$($-k_x, k_y, -k_z$), $\Omega_{zx}$($-k_x, k_y, -k_z$), -$\Omega_{xy}$($-k_x, k_y, -k_z$)]. When $m_z$ is taken into consideration, [$\Omega_{yz}$($-k_x, k_y, k_z$), $-\Omega_{zx}$($-k_x, k_y, k_z$), $-\Omega_{xy}$($-k_x, k_y, k_z$)] are derived. Finally, when considering the time-reversal operation, [$-\Omega_{yz}$($k_x, -k_y, -k_z$), $\Omega_{zx}$($k_x, -k_y, -k_z$), $\Omega_{xy}$($k_x, -k_y, -k_z$)] are obtained. Consequently, the Hall vector \textbf{\textit{h}} lacks an $x$ component and thus takes the form (0, -$\sigma_{xz}$, $\sigma_{xy}$), which is consistent with previous symmetry analysis results\cite{2015symmetry} and double-checked using the \textit{Symmetr} software\cite{website}.
As shown in Fig. 3(a), the whole AHC tensor 
elements transition from zero to a maximum at a rotation 
angle of 45 degrees and ultimately return to zero as the 
N\'eel vector rotates, excluding $\sigma_{xy}$. Our calculations 
indicate that the strongest AHC, $\sigma_{xz} \approx$ 72 S/cm, 
occurs when the N\'eel vector is rotated 45 degrees from 
\textbf{\textit{c}} to $\frac{1}{2}$\textbf{\textit{a}}+\textbf{\textit{b}}. Figure 3(b) 
shows its dependence on energy, revealing a sharp peak of 
approximately 98 S/cm located to the left of the Fermi level, 
suggesting that a rapid increase in AHC can be expected 
with light hole doping.

We next examine the dependence of ANC on the orientation 
of the N\'eel vector. Table II presents the calculated values 
of the nonzero elements of the ANC tensor as the N\'eel 
vector is rotated. Notably, the N\'eel vector 
corresponding to the strongest AHC exhibits a maximal 
ANC value of -0.19 A/m/K. Figure 4(a) presents its 
dependence on energy. At low temperatures, ANC can be 
derived from AHC using Mott relation
\begin{equation}
    \alpha_{i j}=\frac{\pi^2}{3} \frac{k_B^2 T}{e} \frac{\partial \sigma_{i j}}{\partial E}\left(E_F\right).
\end{equation}
The pink dashed line in Fig. 4(a) represents the ANC 
at $T$ = 50 K derived from the AHC shown in Fig. 3(b) 
using Eq. (4). The large value of ANC around the Fermi 
level arises from the sharp variation of AHC at the 
corresponding energy position. The ANC calculated 
using the BC formalism (see the blue 
solid line in Fig. 4(a) using Eq. (3)) agrees well 
with the results from the Mott relation.

From an application perspective, investigating the 
dependence of the ANC on temperature is meaningful. 
Given that CrSb has a high Néel temperature, we 
calculated the ANC at temperatures around room 
temperature. Figure 4(b) shows the calculated ANC 
at the Fermi level as a function of temperature. 
The absolute value of ANC increases sharply from 
50 K to 125 K, then decreases gradually 
from 125 K to 450 K. The maximum absolute value of 
ANC is about 0.224 A/m/K, which is lower than 
that of typical ferromagnets\cite{ANE-npj-database} 
but comparable to RuO$_2$\cite{crystal-thermal-transport}. 
Notably, the ANC remains significant at -0.185 A/m/K
at room temperature, and its decrease slows as the 
temperature rises, indicating potential for thermoelectric 
applications.

\subsection{Berry curvature distribution}
To gain deeper insight into the significant AHC and ANC with the N\'eel vector oriented at 45 degrees from $\textbf{\textit{c}}$ during its rotation from $\textbf{\textit{c}}$ to $\frac{1}{2}$\textbf{\textit{a}}+\textbf{\textit{b}}, we investigate the origin of them by analyzing the BC distribution within the Brillouin zone. Figure 5(a) shows the band structure along 
the $M-C-B$ path, excluding SOC, where $C$ is located at the 
midpoint of $A$-$\Gamma$ and $B$ at the midpoint of $M$-$\Gamma$, 
as shown in Fig. 1(b). The spin-down bands exhibit two crossing 
points near the Fermi level, indicated by green circles. As shown
in Fig. 5(b), when SOC is included, these crossing points open gaps, 
positioning the Fermi level within the gap along the $C-B$ path, 
which contributes to a nonzero BC at the Fermi level. 
A thorough analysis of the crossing points across the entire 
Brillouin zone reveals multiple nodal rings formed by both spin-up 
and spin-down bands. Figure 5(c) shows the locations of 
these nodal rings, with red and blue lines representing those 
formed by spin-up and spin-down bands, respectively. The nodal 
rings can be categorized into two groups for each spin. One group consists of two nodal rings lying in the $M-C-B$ 
plane, with three equivalent planes in the Brillouin zone, 
resulting in a total of six nodal rings protected by 
corresponding mirror planes. The other group comprises 
a wave-like nodal ring around the $k_z = 0$ plane, protected 
by a complex conjugation symmetry, as discussed in reference 
\cite{CrSb-fermi-arc}. Nodal rings with different spin 
characters can be related by the $M_z$ mirror operation. 
With the inclusion of SOC, these nodal rings become gapped, 
resulting in significant BC around the former nodal rings. 
Figure 5(d) displays the BC distribution and 
the nodal rings, showing intimate connection between them. 
Thus, we conclude that the significant AHC and ANC in CrSb 
arise from the SOC-induced gap openings of the numerous 
nodal rings.

\subsection{Experimental setups for detecting the anomalous Hall effect}
At last, we related our theoretical results to the practical realization of such configurations in experiments. Park $et ~al.$ reported that the N\'eel vector of an antiferromagnetic IrMn film can be tuned by an adjacent ferromagnetic film, where exchange coupling between the ferromagnet and antiferromagnet is the driving force \cite{park2011}. Based on this experimental finding, we designed experimental setups to achieve the aforementioned rotations of the N\'eel vector from $\textbf{\textit{c}}$ to desired orientations and to detect the anomalous Hall signal. For example, Fig. 6(a) presents the experimental setup for the detection of $\sigma_{xz}$ when the N\'eel vector rotates from $\textbf{\textit{c}}$ to $\textbf{\textit{a}}$. The CrSb (01$\bar{1}$0) film grown along $\frac{1}{2}$\textbf{\textit{a}}+\textbf{\textit{b}} is adequate for measuring the anomalous Hall signal originating from nonzero $\sigma_{xz}$. The orientation of the N\'eel vector along $\textbf{\textit{c}}$ in the ground state would tilt towards the direction of the ferromagnetic moments in the ferromagnetic substrate. The canting of the N\'eel vector could be less prominent than the rotation due to the strong antiferromagnetic coupling between opposite magnetic sublattices in CrSb, which gives rise to the rather high N\'eel temperature (exceeding 700 K). In fact, although N\'eel vector canting could be induced by damping-like spin-orbit torque and Dzyaloshinskii-Moriya torque, the observed AHE signal has been demonstrated to originate primarily from the altermagnetic order in the mirror-symmetry-broken CrSb films rather than from canting \cite{CrSb_nature}. Note that the magnetic moments of the ferromagnetic substrate in Fig. 6(a) are along $\textbf{\textit{a}}$, and thus the ferromagnetic substrate does not contribute to the anomalous Hall signal in this setup, owing to the common expression $j^{\text{AHE}} \sim M \times E$, which is suitable for most ferromagnets. As shown by the gray line in Fig. 3(a), once the N\'eel vector tilts from $\textbf{\textit{c}}$ to $\textbf{\textit{a}}$, $\sigma_{xz}$ becomes nonzero, the AHE signal can be detected. The ease and amplitude of N\'eel vector rotation driven by the ferromagnetic substrate depend on external stimuli such as temperature, the exchange energy between the ferromagnetic substrate and CrSb, and the antiferromagnetic exchange stiffness in CrSb.

Figure 6(b) presents the experimental setup for detecting $\sigma_{xz}$ when the N\'eel vector rotates from $\textbf{\textit{c}}$ to $\frac{1}{2}$\textbf{\textit{a}}+\textbf{\textit{b}}, as shown by the purple line in Fig. 3(a). To induce the rotation of the N\'eel vector towards $\frac{1}{2}$\textbf{\textit{a}}+\textbf{\textit{b}}, the ferromagnetic substrate should have an easy-axis magnetocrystalline anisotropy along $\frac{1}{2}$\textbf{\textit{a}}+\textbf{\textit{b}}. Most importantly, the ferromagnetic substrate should preferably be an insulator, as metallic substrates with magnetic moments along $\frac{1}{2}$\textbf{\textit{a}}+\textbf{\textit{b}} would contribute to an additional AHE signal, thereby hindering the detection of the intrinsic AHE signal from CrSb. Therefore, insulating ferromagnetic materials, such as CrI$_3$ and Cr$_2$Ge$_2$Te$_6$, are preferred for the substrate.

\section{Discussion}
In summary, our DFT calculations revealed the magnetism 
and anomalous transport properties of CrSb. The weak MAE suggests flexibility in adjusting the N\'eel vector orientation. By examining the dependence of the AHC and ANC on the orientation of the N\'eel vector, we determined that alignment along $\frac{1}{2}$\textbf{\textit{a}}+\textbf{\textit{b}} results in significant values for both AHC and ANC. These pronounced values are attributed to SOC-induced gap openings at numerous nodal rings. The substantial AHC and ANC, together with other intriguing features such as the high N\'eel temperature, surface Fermi arcs, and high-mobility carriers, establish CrSb as a compelling platform for investigating novel altermagnetism-related properties.

\section*{Data Availability}
Data are available from the corresponding authors upon reasonable request.

\begin{acknowledgments}
This work was supported by the National Key R\&D Program 
of China (Grant No. 2021YFB3501503), the National 
Natural Science Foundation of China (Grants No. 52271016 and
No. 52188101), Chinese Academy of 
Sciences Project for Young Scientists in Basic 
Research Grant No.YSBR-109, and Foundation from 
Liaoning Province (Grant No. XLYC2203080).
\end{acknowledgments}

\section*{Author Contributions}
D.S., X.C. and Y.S. conceived and designed the research project. T.Y. and S.I. performed the calculations. T.Y., I.S., P.L., D.S., X.C. and Y.S. wrote the paper.

\section*{Competing Interests}
The authors declare no competing interests.



\begin{thebibliography}{99}%
\makeatletter
\providecommand \@ifxundefined [1]{%
 \@ifx{#1\undefined}
}%
\providecommand \@ifnum [1]{%
 \ifnum #1\expandafter \@firstoftwo
 \else \expandafter \@secondoftwo
 \fi
}%
\providecommand \@ifx [1]{%
 \ifx #1\expandafter \@firstoftwo
 \else \expandafter \@secondoftwo
 \fi
}%
\providecommand \natexlab [1]{#1}%
\providecommand \enquote  [1]{``#1''}%
\providecommand \bibnamefont  [1]{#1}%
\providecommand \bibfnamefont [1]{#1}%
\providecommand \citenamefont [1]{#1}%
\providecommand \href@noop [0]{\@secondoftwo}%
\providecommand \href [0]{\begingroup \@sanitize@url \@href}%
\providecommand \@href[1]{\@@startlink{#1}\@@href}%
\providecommand \@@href[1]{\endgroup#1\@@endlink}%
\providecommand \@sanitize@url [0]{\catcode `\\12\catcode `\$12\catcode `\&12\catcode `\#12\catcode `\^12\catcode `\_12\catcode `\%12\relax}%
\providecommand \@@startlink[1]{}%
\providecommand \@@endlink[0]{}%
\providecommand \url  [0]{\begingroup\@sanitize@url \@url }%
\providecommand \@url [1]{\endgroup\@href {#1}{\urlprefix }}%
\providecommand \urlprefix  [0]{URL }%
\providecommand \Eprint [0]{\href }%
\providecommand \doibase [0]{https://doi.org/}%
\providecommand \selectlanguage [0]{\@gobble}%
\providecommand \bibinfo  [0]{\@secondoftwo}%
\providecommand \bibfield  [0]{\@secondoftwo}%
\providecommand \translation [1]{[#1]}%
\providecommand \BibitemOpen [0]{}%
\providecommand \bibitemStop [0]{}%
\providecommand \bibitemNoStop [0]{.\EOS\space}%
\providecommand \EOS [0]{\spacefactor3000\relax}%
\providecommand \BibitemShut  [1]{\csname bibitem#1\endcsname}%
\let\auto@bib@innerbib\@empty
\bibitem [{\citenamefont {{\v{S}}mejkal}\ \emph {et~al.}(2022{\natexlab{a}})\citenamefont {{\v{S}}mejkal}, \citenamefont {MacDonald}, \citenamefont {Sinova}, \citenamefont {Nakatsuji},\ and\ \citenamefont {Jungwirth}}]{altermagnetism1}%
  \BibitemOpen
  \bibfield  {author} {\bibinfo {author} {\bibfnamefont {L.}~\bibnamefont {{\v{S}}mejkal}}, \bibinfo {author} {\bibfnamefont {A.~H.}\ \bibnamefont {MacDonald}}, \bibinfo {author} {\bibfnamefont {J.}~\bibnamefont {Sinova}}, \bibinfo {author} {\bibfnamefont {S.}~\bibnamefont {Nakatsuji}},\ and\ \bibinfo {author} {\bibfnamefont {T.}~\bibnamefont {Jungwirth}},\ }\bibfield  {title} {\bibinfo {title} {Anomalous Hall antiferromagnets},\ }\href@noop {} {\bibfield  {journal} {\bibinfo  {journal} {Nature Reviews Materials}\ }\textbf {\bibinfo {volume} {7}},\ \bibinfo {pages} {482} (\bibinfo {year} {2022}{\natexlab{a}})}\BibitemShut {NoStop}%
\bibitem [{\citenamefont {{\v{S}}mejkal}\ \emph {et~al.}(2022{\natexlab{b}})\citenamefont {{\v{S}}mejkal}, \citenamefont {Sinova},\ and\ \citenamefont {Jungwirth}}]{altermagnetism2}%
  \BibitemOpen
  \bibfield  {author} {\bibinfo {author} {\bibfnamefont {L.}~\bibnamefont {{\v{S}}mejkal}}, \bibinfo {author} {\bibfnamefont {J.}~\bibnamefont {Sinova}},\ and\ \bibinfo {author} {\bibfnamefont {T.}~\bibnamefont {Jungwirth}},\ }\bibfield  {title} {\bibinfo {title} {Beyond conventional ferromagnetism and antiferromagnetism: A phase with nonrelativistic spin and crystal rotation symmetry},\ }\href@noop {} {\bibfield  {journal} {\bibinfo  {journal} {Physical Review X}\ }\textbf {\bibinfo {volume} {12}},\ \bibinfo {pages} {031042} (\bibinfo {year} {2022}{\natexlab{b}})}\BibitemShut {NoStop}%
\bibitem [{\citenamefont {Mazin}\ and\ \citenamefont {Editors}(2022)}]{altermagnetism3}%
  \BibitemOpen
  \bibfield  {author} {\bibinfo {author} {\bibfnamefont {I.}~\bibnamefont {Mazin}}\ and\ \bibinfo {author} {\bibfnamefont {P.}~\bibnamefont {Editors}},\ }\href@noop {} {\bibinfo {title} {Altermagnetism—a new punch line of fundamental magnetism}} (\bibinfo {year} {2022})\BibitemShut {NoStop}%
\bibitem [{\citenamefont {{\v{S}}mejkal}\ \emph {et~al.}(2022{\natexlab{c}})\citenamefont {{\v{S}}mejkal}, \citenamefont {Sinova},\ and\ \citenamefont {Jungwirth}}]{altermagnetism4}%
  \BibitemOpen
  \bibfield  {author} {\bibinfo {author} {\bibfnamefont {L.}~\bibnamefont {{\v{S}}mejkal}}, \bibinfo {author} {\bibfnamefont {J.}~\bibnamefont {Sinova}},\ and\ \bibinfo {author} {\bibfnamefont {T.}~\bibnamefont {Jungwirth}},\ }\bibfield  {title} {\bibinfo {title} {Emerging research landscape of altermagnetism},\ }\href@noop {} {\bibfield  {journal} {\bibinfo  {journal} {Physical Review X}\ }\textbf {\bibinfo {volume} {12}},\ \bibinfo {pages} {040501} (\bibinfo {year} {2022}{\natexlab{c}})}\BibitemShut {NoStop}%
\bibitem [{\citenamefont {Mazin}(2024)}]{altermagnetism5}%
  \BibitemOpen
  \bibfield  {author} {\bibinfo {author} {\bibfnamefont {I.}~\bibnamefont {Mazin}},\ }\bibfield  {title} {\bibinfo {title} {Altermagnetism then and now},\ }\href@noop {} {\bibfield  {journal} {\bibinfo  {journal} {Physics}\ }\textbf {\bibinfo {volume} {17}},\ \bibinfo {pages} {4} (\bibinfo {year} {2024})}\BibitemShut {NoStop}%
\bibitem [{\citenamefont {Yan}\ \emph {et~al.}(2024)\citenamefont {Yan}, \citenamefont {Zhou}, \citenamefont {Qin},\ and\ \citenamefont {Liu}}]{altermagnetism6}%
  \BibitemOpen
  \bibfield  {author} {\bibinfo {author} {\bibfnamefont {H.}~\bibnamefont {Yan}}, \bibinfo {author} {\bibfnamefont {X.}~\bibnamefont {Zhou}}, \bibinfo {author} {\bibfnamefont {P.}~\bibnamefont {Qin}},\ and\ \bibinfo {author} {\bibfnamefont {Z.}~\bibnamefont {Liu}},\ }\bibfield  {title} {\bibinfo {title} {Review on spin-split antiferromagnetic spintronics},\ }\href@noop {} {\bibfield  {journal} {\bibinfo  {journal} {Applied Physics Letters}\ }\textbf {\bibinfo {volume} {124}} (\bibinfo {year} {2024})}\BibitemShut {NoStop}%
\bibitem [{\citenamefont {Bai}\ \emph {et~al.}(2024)\citenamefont {Bai}, \citenamefont {Feng}, \citenamefont {Liu}, \citenamefont {{\v{S}}mejkal}, \citenamefont {Mokrousov},\ and\ \citenamefont {Yao}}]{altermagnetism7}%
  \BibitemOpen
  \bibfield  {author} {\bibinfo {author} {\bibfnamefont {L.}~\bibnamefont {Bai}}, \bibinfo {author} {\bibfnamefont {W.}~\bibnamefont {Feng}}, \bibinfo {author} {\bibfnamefont {S.}~\bibnamefont {Liu}}, \bibinfo {author} {\bibfnamefont {L.}~\bibnamefont {{\v{S}}mejkal}}, \bibinfo {author} {\bibfnamefont {Y.}~\bibnamefont {Mokrousov}},\ and\ \bibinfo {author} {\bibfnamefont {Y.}~\bibnamefont {Yao}},\ }\bibfield  {title} {\bibinfo {title} {Altermagnetism: Exploring new frontiers in magnetism and spintronics},\ }\href@noop {} {\bibfield  {journal} {\bibinfo  {journal} {Advanced Functional Materials}\ ,\ \bibinfo {pages} {2409327}} (\bibinfo {year} {2024})}\BibitemShut {NoStop}%
\bibitem [{\citenamefont {{\v{S}}mejkal}\ \emph {et~al.}(2020)\citenamefont {{\v{S}}mejkal}, \citenamefont {Gonz{\'a}lez-Hern{\'a}ndez}, \citenamefont {Jungwirth},\ and\ \citenamefont {Sinova}}]{science_advance_RuO2}%
  \BibitemOpen
  \bibfield  {author} {\bibinfo {author} {\bibfnamefont {L.}~\bibnamefont {{\v{S}}mejkal}}, \bibinfo {author} {\bibfnamefont {R.}~\bibnamefont {Gonz{\'a}lez-Hern{\'a}ndez}}, \bibinfo {author} {\bibfnamefont {T.}~\bibnamefont {Jungwirth}},\ and\ \bibinfo {author} {\bibfnamefont {J.}~\bibnamefont {Sinova}},\ }\bibfield  {title} {\bibinfo {title} {Crystal time-reversal symmetry breaking and spontaneous Hall effect in collinear antiferromagnets},\ }\href@noop {} {\bibfield  {journal} {\bibinfo  {journal} {Science advances}\ }\textbf {\bibinfo {volume} {6}},\ \bibinfo {pages} {eaaz8809} (\bibinfo {year} {2020})}\BibitemShut {NoStop}%
\bibitem [{\citenamefont {Mazin}\ \emph {et~al.}(2021)\citenamefont {Mazin}, \citenamefont {Koepernik}, \citenamefont {Johannes}, \citenamefont {Gonz{\'a}lez-Hern{\'a}ndez},\ and\ \citenamefont {{\v{S}}mejkal}}]{FeSb2}%
  \BibitemOpen
  \bibfield  {author} {\bibinfo {author} {\bibfnamefont {I.~I.}\ \bibnamefont {Mazin}}, \bibinfo {author} {\bibfnamefont {K.}~\bibnamefont {Koepernik}}, \bibinfo {author} {\bibfnamefont {M.~D.}\ \bibnamefont {Johannes}}, \bibinfo {author} {\bibfnamefont {R.}~\bibnamefont {Gonz{\'a}lez-Hern{\'a}ndez}},\ and\ \bibinfo {author} {\bibfnamefont {L.}~\bibnamefont {{\v{S}}mejkal}},\ }\bibfield  {title} {\bibinfo {title} {Prediction of unconventional magnetism in doped fesb2},\ }\href@noop {} {\bibfield  {journal} {\bibinfo  {journal} {Proceedings of the National Academy of Sciences}\ }\textbf {\bibinfo {volume} {118}},\ \bibinfo {pages} {e2108924118} (\bibinfo {year} {2021})}\BibitemShut {NoStop}%
\bibitem [{\citenamefont {Ma}\ \emph {et~al.}(2021)\citenamefont {Ma}, \citenamefont {Hu}, \citenamefont {Li}, \citenamefont {Liu}, \citenamefont {Yao}, \citenamefont {Jia},\ and\ \citenamefont {Liu}}]{liu2021}%
  \BibitemOpen
  \bibfield  {author} {\bibinfo {author} {\bibfnamefont {H.-Y.}\ \bibnamefont {Ma}}, \bibinfo {author} {\bibfnamefont {M.}~\bibnamefont {Hu}}, \bibinfo {author} {\bibfnamefont {N.}~\bibnamefont {Li}}, \bibinfo {author} {\bibfnamefont {J.}~\bibnamefont {Liu}}, \bibinfo {author} {\bibfnamefont {W.}~\bibnamefont {Yao}}, \bibinfo {author} {\bibfnamefont {J.-F.}\ \bibnamefont {Jia}},\ and\ \bibinfo {author} {\bibfnamefont {J.}~\bibnamefont {Liu}},\ }\bibfield  {title} {\bibinfo {title} {Multifunctional antiferromagnetic materials with giant piezomagnetism and noncollinear spin current},\ }\href@noop {} {\bibfield  {journal} {\bibinfo  {journal} {Nature Communications}\ }\textbf {\bibinfo {volume} {12}},\ \bibinfo {pages} {2846} (\bibinfo {year} {2021})}\BibitemShut {NoStop}%
\bibitem [{\citenamefont {Del~Re}(2024)}]{arxive-correlated-altermagnets}%
  \BibitemOpen
  \bibfield  {author} {\bibinfo {author} {\bibfnamefont {L.}~\bibnamefont {Del~Re}},\ }\bibfield  {title} {\bibinfo {title} {Dirac points and topological phases in correlated altermagnets},\ }\href@noop {} {\bibfield  {journal} {\bibinfo  {journal} {arXiv preprint arXiv:2408.14288}\ } (\bibinfo {year} {2024})}\BibitemShut {NoStop}%
\bibitem [{\citenamefont {Zhou}\ \emph {et~al.}(2024)\citenamefont {Zhou}, \citenamefont {Feng}, \citenamefont {Zhang}, \citenamefont {{\v{S}}mejkal}, \citenamefont {Sinova}, \citenamefont {Mokrousov},\ and\ \citenamefont {Yao}}]{crystal-thermal-transport}%
  \BibitemOpen
  \bibfield  {author} {\bibinfo {author} {\bibfnamefont {X.}~\bibnamefont {Zhou}}, \bibinfo {author} {\bibfnamefont {W.}~\bibnamefont {Feng}}, \bibinfo {author} {\bibfnamefont {R.-W.}\ \bibnamefont {Zhang}}, \bibinfo {author} {\bibfnamefont {L.}~\bibnamefont {{\v{S}}mejkal}}, \bibinfo {author} {\bibfnamefont {J.}~\bibnamefont {Sinova}}, \bibinfo {author} {\bibfnamefont {Y.}~\bibnamefont {Mokrousov}},\ and\ \bibinfo {author} {\bibfnamefont {Y.}~\bibnamefont {Yao}},\ }\bibfield  {title} {\bibinfo {title} {Crystal thermal transport in altermagnetic RuO$_2$},\ }\href@noop {} {\bibfield  {journal} {\bibinfo  {journal} {Physical review letters}\ }\textbf {\bibinfo {volume} {132}},\ \bibinfo {pages} {056701} (\bibinfo {year} {2024})}\BibitemShut {NoStop}%
\bibitem [{\citenamefont {Nguyen}\ and\ \citenamefont {Yamauchi}(2023)}]{DFT-CaCrO3-AHC}%
  \BibitemOpen
  \bibfield  {author} {\bibinfo {author} {\bibfnamefont {T.~P.~T.}\ \bibnamefont {Nguyen}}\ and\ \bibinfo {author} {\bibfnamefont {K.}~\bibnamefont {Yamauchi}},\ }\bibfield  {title} {\bibinfo {title} {Ab initio prediction of anomalous Hall effect in antiferromagnetic CaCrO$_3$},\ }\href@noop {} {\bibfield  {journal} {\bibinfo  {journal} {Physical Review B}\ }\textbf {\bibinfo {volume} {107}},\ \bibinfo {pages} {155126} (\bibinfo {year} {2023})}\BibitemShut {NoStop}%
\bibitem [{\citenamefont {Chakraborty}\ \emph {et~al.}(2024)\citenamefont {Chakraborty}, \citenamefont {Gonz{\'a}lez~Hern{\'a}ndez}, \citenamefont {{\v{S}}mejkal},\ and\ \citenamefont {Sinova}}]{DFT-strain-ReO2}%
  \BibitemOpen
  \bibfield  {author} {\bibinfo {author} {\bibfnamefont {A.}~\bibnamefont {Chakraborty}}, \bibinfo {author} {\bibfnamefont {R.}~\bibnamefont {Gonz{\'a}lez~Hern{\'a}ndez}}, \bibinfo {author} {\bibfnamefont {L.}~\bibnamefont {{\v{S}}mejkal}},\ and\ \bibinfo {author} {\bibfnamefont {J.}~\bibnamefont {Sinova}},\ }\bibfield  {title} {\bibinfo {title} {Strain-induced phase transition from antiferromagnet to altermagnet},\ }\href@noop {} {\bibfield  {journal} {\bibinfo  {journal} {Physical Review B}\ }\textbf {\bibinfo {volume} {109}},\ \bibinfo {pages} {144421} (\bibinfo {year} {2024})}\BibitemShut {NoStop}%
\bibitem [{\citenamefont {Guo}\ \emph {et~al.}(2023{\natexlab{a}})\citenamefont {Guo}, \citenamefont {Liu}, \citenamefont {Janson}, \citenamefont {Fulga}, \citenamefont {van~den Brink},\ and\ \citenamefont {Facio}}]{high_througput_spin_splitting_DFT_calculations}%
  \BibitemOpen
  \bibfield  {author} {\bibinfo {author} {\bibfnamefont {Y.}~\bibnamefont {Guo}}, \bibinfo {author} {\bibfnamefont {H.}~\bibnamefont {Liu}}, \bibinfo {author} {\bibfnamefont {O.}~\bibnamefont {Janson}}, \bibinfo {author} {\bibfnamefont {I.~C.}\ \bibnamefont {Fulga}}, \bibinfo {author} {\bibfnamefont {J.}~\bibnamefont {van~den Brink}},\ and\ \bibinfo {author} {\bibfnamefont {J.~I.}\ \bibnamefont {Facio}},\ }\bibfield  {title} {\bibinfo {title} {Spin-split collinear antiferromagnets: A large-scale ab-initio study},\ }\href@noop {} {\bibfield  {journal} {\bibinfo  {journal} {Materials Today Physics}\ }\textbf {\bibinfo {volume} {32}},\ \bibinfo {pages} {100991} (\bibinfo {year} {2023}{\natexlab{a}})}\BibitemShut {NoStop}%
\bibitem [{\citenamefont {{\v{S}}mejkal}\ \emph {et~al.}(2023)\citenamefont {{\v{S}}mejkal}, \citenamefont {Marmodoro}, \citenamefont {Ahn}, \citenamefont {Gonz{\'a}lez-Hern{\'a}ndez}, \citenamefont {Turek}, \citenamefont {Mankovsky}, \citenamefont {Ebert}, \citenamefont {D’Souza}, \citenamefont {{\v{S}}ipr}, \citenamefont {Sinova} \emph {et~al.}}]{magnon-RuO2-PRL}%
  \BibitemOpen
  \bibfield  {author} {\bibinfo {author} {\bibfnamefont {L.}~\bibnamefont {{\v{S}}mejkal}}, \bibinfo {author} {\bibfnamefont {A.}~\bibnamefont {Marmodoro}}, \bibinfo {author} {\bibfnamefont {K.-H.}\ \bibnamefont {Ahn}}, \bibinfo {author} {\bibfnamefont {R.}~\bibnamefont {Gonz{\'a}lez-Hern{\'a}ndez}}, \bibinfo {author} {\bibfnamefont {I.}~\bibnamefont {Turek}}, \bibinfo {author} {\bibfnamefont {S.}~\bibnamefont {Mankovsky}}, \bibinfo {author} {\bibfnamefont {H.}~\bibnamefont {Ebert}}, \bibinfo {author} {\bibfnamefont {S.~W.}\ \bibnamefont {D’Souza}}, \bibinfo {author} {\bibfnamefont {O.}~\bibnamefont {{\v{S}}ipr}}, \bibinfo {author} {\bibfnamefont {J.}~\bibnamefont {Sinova}}, \emph {et~al.},\ }\bibfield  {title} {\bibinfo {title} {Chiral magnons in altermagnetic RuO$_2$},\ }\href@noop {} {\bibfield  {journal} {\bibinfo  {journal} {Physical Review Letters}\ }\textbf {\bibinfo {volume} {131}},\ \bibinfo {pages} {256703} (\bibinfo {year} {2023})}\BibitemShut {NoStop}%
\bibitem [{\citenamefont {Roig}\ \emph {et~al.}(2024)\citenamefont {Roig}, \citenamefont {Kreisel}, \citenamefont {Yu}, \citenamefont {Andersen},\ and\ \citenamefont {Agterberg}}]{minimal-models}%
  \BibitemOpen
  \bibfield  {author} {\bibinfo {author} {\bibfnamefont {M.}~\bibnamefont {Roig}}, \bibinfo {author} {\bibfnamefont {A.}~\bibnamefont {Kreisel}}, \bibinfo {author} {\bibfnamefont {Y.}~\bibnamefont {Yu}}, \bibinfo {author} {\bibfnamefont {B.~M.}\ \bibnamefont {Andersen}},\ and\ \bibinfo {author} {\bibfnamefont {D.~F.}\ \bibnamefont {Agterberg}},\ }\bibfield  {title} {\bibinfo {title} {Minimal models for altermagnetism},\ }\href@noop {} {\bibfield  {journal} {\bibinfo  {journal} {Physical Review B}\ }\textbf {\bibinfo {volume} {110}},\ \bibinfo {pages} {144412} (\bibinfo {year} {2024})}\BibitemShut {NoStop}%
\bibitem [{\citenamefont {Zhou}\ \emph {et~al.}(2021)\citenamefont {Zhou}, \citenamefont {Feng}, \citenamefont {Yang}, \citenamefont {Guo},\ and\ \citenamefont {Yao}}]{moke_RuO2}%
  \BibitemOpen
  \bibfield  {author} {\bibinfo {author} {\bibfnamefont {X.}~\bibnamefont {Zhou}}, \bibinfo {author} {\bibfnamefont {W.}~\bibnamefont {Feng}}, \bibinfo {author} {\bibfnamefont {X.}~\bibnamefont {Yang}}, \bibinfo {author} {\bibfnamefont {G.-Y.}\ \bibnamefont {Guo}},\ and\ \bibinfo {author} {\bibfnamefont {Y.}~\bibnamefont {Yao}},\ }\bibfield  {title} {\bibinfo {title} {Crystal chirality magneto-optical effects in collinear antiferromagnets},\ }\href@noop {} {\bibfield  {journal} {\bibinfo  {journal} {Physical Review B}\ }\textbf {\bibinfo {volume} {104}},\ \bibinfo {pages} {024401} (\bibinfo {year} {2021})}\BibitemShut {NoStop}%
\bibitem [{\citenamefont {Cuono}\ \emph {et~al.}(2023)\citenamefont {Cuono}, \citenamefont {Sattigeri}, \citenamefont {Skolimowski},\ and\ \citenamefont {Autieri}}]{Orbital-selective-and-correlation-altermagnetism}%
  \BibitemOpen
  \bibfield  {author} {\bibinfo {author} {\bibfnamefont {G.}~\bibnamefont {Cuono}}, \bibinfo {author} {\bibfnamefont {R.~M.}\ \bibnamefont {Sattigeri}}, \bibinfo {author} {\bibfnamefont {J.}~\bibnamefont {Skolimowski}},\ and\ \bibinfo {author} {\bibfnamefont {C.}~\bibnamefont {Autieri}},\ }\bibfield  {title} {\bibinfo {title} {Orbital-selective altermagnetism and correlation-enhanced spin-splitting in strongly-correlated transition metal oxides},\ }\href@noop {} {\bibfield  {journal} {\bibinfo  {journal} {Journal of Magnetism and Magnetic Materials}\ }\textbf {\bibinfo {volume} {586}},\ \bibinfo {pages} {171163} (\bibinfo {year} {2023})}\BibitemShut {NoStop}%
\bibitem [{\citenamefont {Yuan}\ \emph {et~al.}(2020)\citenamefont {Yuan}, \citenamefont {Wang}, \citenamefont {Luo}, \citenamefont {Rashba},\ and\ \citenamefont {Zunger}}]{PRB-Tables2}%
  \BibitemOpen
  \bibfield  {author} {\bibinfo {author} {\bibfnamefont {L.-D.}\ \bibnamefont {Yuan}}, \bibinfo {author} {\bibfnamefont {Z.}~\bibnamefont {Wang}}, \bibinfo {author} {\bibfnamefont {J.-W.}\ \bibnamefont {Luo}}, \bibinfo {author} {\bibfnamefont {E.~I.}\ \bibnamefont {Rashba}},\ and\ \bibinfo {author} {\bibfnamefont {A.}~\bibnamefont {Zunger}},\ }\bibfield  {title} {\bibinfo {title} {Giant momentum-dependent spin splitting in centrosymmetric low-z antiferromagnets},\ }\href@noop {} {\bibfield  {journal} {\bibinfo  {journal} {Physical Review B}\ }\textbf {\bibinfo {volume} {102}},\ \bibinfo {pages} {014422} (\bibinfo {year} {2020})}\BibitemShut {NoStop}%
\bibitem [{\citenamefont {Yuan}\ \emph {et~al.}(2021)\citenamefont {Yuan}, \citenamefont {Wang}, \citenamefont {Luo},\ and\ \citenamefont {Zunger}}]{PRM-tables}%
  \BibitemOpen
  \bibfield  {author} {\bibinfo {author} {\bibfnamefont {L.-D.}\ \bibnamefont {Yuan}}, \bibinfo {author} {\bibfnamefont {Z.}~\bibnamefont {Wang}}, \bibinfo {author} {\bibfnamefont {J.-W.}\ \bibnamefont {Luo}},\ and\ \bibinfo {author} {\bibfnamefont {A.}~\bibnamefont {Zunger}},\ }\bibfield  {title} {\bibinfo {title} {Prediction of low-z collinear and noncollinear antiferromagnetic compounds having momentum-dependent spin splitting even without spin-orbit coupling},\ }\href@noop {} {\bibfield  {journal} {\bibinfo  {journal} {Physical Review Materials}\ }\textbf {\bibinfo {volume} {5}},\ \bibinfo {pages} {014409} (\bibinfo {year} {2021})}\BibitemShut {NoStop}%
\bibitem [{\citenamefont {Fakhredine}\ \emph {et~al.}(2023)\citenamefont {Fakhredine}, \citenamefont {Sattigeri}, \citenamefont {Cuono},\ and\ \citenamefont {Autieri}}]{space_group_62}%
  \BibitemOpen
  \bibfield  {author} {\bibinfo {author} {\bibfnamefont {A.}~\bibnamefont {Fakhredine}}, \bibinfo {author} {\bibfnamefont {R.~M.}\ \bibnamefont {Sattigeri}}, \bibinfo {author} {\bibfnamefont {G.}~\bibnamefont {Cuono}},\ and\ \bibinfo {author} {\bibfnamefont {C.}~\bibnamefont {Autieri}},\ }\bibfield  {title} {\bibinfo {title} {Interplay between altermagnetism and nonsymmorphic symmetries generating large anomalous Hall conductivity by semi-dirac points induced anticrossings},\ }\href@noop {} {\bibfield  {journal} {\bibinfo  {journal} {Physical Review B}\ }\textbf {\bibinfo {volume} {108}},\ \bibinfo {pages} {115138} (\bibinfo {year} {2023})}\BibitemShut {NoStop}%
\bibitem [{\citenamefont {Zhu}\ \emph {et~al.}(2023)\citenamefont {Zhu}, \citenamefont {Zhuang}, \citenamefont {Wu},\ and\ \citenamefont {Yan}}]{topological-superconductivity1}%
  \BibitemOpen
  \bibfield  {author} {\bibinfo {author} {\bibfnamefont {D.}~\bibnamefont {Zhu}}, \bibinfo {author} {\bibfnamefont {Z.-Y.}\ \bibnamefont {Zhuang}}, \bibinfo {author} {\bibfnamefont {Z.}~\bibnamefont {Wu}},\ and\ \bibinfo {author} {\bibfnamefont {Z.}~\bibnamefont {Yan}},\ }\bibfield  {title} {\bibinfo {title} {Topological superconductivity in two-dimensional altermagnetic metals},\ }\href@noop {} {\bibfield  {journal} {\bibinfo  {journal} {Physical Review B}\ }\textbf {\bibinfo {volume} {108}},\ \bibinfo {pages} {184505} (\bibinfo {year} {2023})}\BibitemShut {NoStop}%
\bibitem [{\citenamefont {Brekke}\ \emph {et~al.}(2023)\citenamefont {Brekke}, \citenamefont {Brataas},\ and\ \citenamefont {Sudb{\o}}}]{topological-superconductivity2}%
  \BibitemOpen
  \bibfield  {author} {\bibinfo {author} {\bibfnamefont {B.}~\bibnamefont {Brekke}}, \bibinfo {author} {\bibfnamefont {A.}~\bibnamefont {Brataas}},\ and\ \bibinfo {author} {\bibfnamefont {A.}~\bibnamefont {Sudb{\o}}},\ }\bibfield  {title} {\bibinfo {title} {Two-dimensional altermagnets: Superconductivity in a minimal microscopic model},\ }\href@noop {} {\bibfield  {journal} {\bibinfo  {journal} {Physical Review B}\ }\textbf {\bibinfo {volume} {108}},\ \bibinfo {pages} {224421} (\bibinfo {year} {2023})}\BibitemShut {NoStop}%
\bibitem [{\citenamefont {Fernandes}\ \emph {et~al.}(2024)\citenamefont {Fernandes}, \citenamefont {De~Carvalho}, \citenamefont {Birol},\ and\ \citenamefont {Pereira}}]{topological-transition}%
  \BibitemOpen
  \bibfield  {author} {\bibinfo {author} {\bibfnamefont {R.~M.}\ \bibnamefont {Fernandes}}, \bibinfo {author} {\bibfnamefont {V.~S.}\ \bibnamefont {De~Carvalho}}, \bibinfo {author} {\bibfnamefont {T.}~\bibnamefont {Birol}},\ and\ \bibinfo {author} {\bibfnamefont {R.~G.}\ \bibnamefont {Pereira}},\ }\bibfield  {title} {\bibinfo {title} {Topological transition from nodal to nodeless zeeman splitting in altermagnets},\ }\href@noop {} {\bibfield  {journal} {\bibinfo  {journal} {Physical Review B}\ }\textbf {\bibinfo {volume} {109}},\ \bibinfo {pages} {024404} (\bibinfo {year} {2024})}\BibitemShut {NoStop}%
\bibitem [{\citenamefont {Zhan}\ \emph {et~al.}(2023)\citenamefont {Zhan}, \citenamefont {Li}, \citenamefont {Shi}, \citenamefont {Chen},\ and\ \citenamefont {Sun}}]{zhanjie_RuO2}%
  \BibitemOpen
  \bibfield  {author} {\bibinfo {author} {\bibfnamefont {J.}~\bibnamefont {Zhan}}, \bibinfo {author} {\bibfnamefont {J.}~\bibnamefont {Li}}, \bibinfo {author} {\bibfnamefont {W.}~\bibnamefont {Shi}}, \bibinfo {author} {\bibfnamefont {X.-Q.}\ \bibnamefont {Chen}},\ and\ \bibinfo {author} {\bibfnamefont {Y.}~\bibnamefont {Sun}},\ }\bibfield  {title} {\bibinfo {title} {Coexistence of weyl semimetal and weyl nodal loop semimetal phases in a collinear antiferromagnet},\ }\href@noop {} {\bibfield  {journal} {\bibinfo  {journal} {Physical Review B}\ }\textbf {\bibinfo {volume} {107}},\ \bibinfo {pages} {224402} (\bibinfo {year} {2023})}\BibitemShut {NoStop}%
\bibitem [{\citenamefont {Hou}\ \emph {et~al.}(2023)\citenamefont {Hou}, \citenamefont {Yang}, \citenamefont {Liu}, \citenamefont {Guo},\ and\ \citenamefont {Lu}}]{zhongyi_LU_Nb2FeB2}%
  \BibitemOpen
  \bibfield  {author} {\bibinfo {author} {\bibfnamefont {X.-Y.}\ \bibnamefont {Hou}}, \bibinfo {author} {\bibfnamefont {H.-C.}\ \bibnamefont {Yang}}, \bibinfo {author} {\bibfnamefont {Z.-X.}\ \bibnamefont {Liu}}, \bibinfo {author} {\bibfnamefont {P.-J.}\ \bibnamefont {Guo}},\ and\ \bibinfo {author} {\bibfnamefont {Z.-Y.}\ \bibnamefont {Lu}},\ }\bibfield  {title} {\bibinfo {title} {Large intrinsic anomalous Hall effect in both Nb$_2$FeB$_2$ and Ta$_2$FeB$_2$ with collinear antiferromagnetism},\ }\href@noop {} {\bibfield  {journal} {\bibinfo  {journal} {Physical Review B}\ }\textbf {\bibinfo {volume} {107}},\ \bibinfo {pages} {L161109} (\bibinfo {year} {2023})}\BibitemShut {NoStop}%
\bibitem [{\citenamefont {Gao}\ \emph {et~al.}(2023)\citenamefont {Gao}, \citenamefont {Qu}, \citenamefont {Zeng}, \citenamefont {Liu}, \citenamefont {Wen}, \citenamefont {Sun}, \citenamefont {Guo},\ and\ \citenamefont {Lu}}]{Zhongyi-Lu-AI}%
  \BibitemOpen
  \bibfield  {author} {\bibinfo {author} {\bibfnamefont {Z.-F.}\ \bibnamefont {Gao}}, \bibinfo {author} {\bibfnamefont {S.}~\bibnamefont {Qu}}, \bibinfo {author} {\bibfnamefont {B.}~\bibnamefont {Zeng}}, \bibinfo {author} {\bibfnamefont {Y.}~\bibnamefont {Liu}}, \bibinfo {author} {\bibfnamefont {J.-R.}\ \bibnamefont {Wen}}, \bibinfo {author} {\bibfnamefont {H.}~\bibnamefont {Sun}}, \bibinfo {author} {\bibfnamefont {P.-J.}\ \bibnamefont {Guo}},\ and\ \bibinfo {author} {\bibfnamefont {Z.-Y.}\ \bibnamefont {Lu}},\ }\bibfield  {title} {\bibinfo {title} {Ai-accelerated discovery of altermagnetic materials},\ }\href@noop {} {\bibfield  {journal} {\bibinfo  {journal} {arXiv preprint arXiv:2311.04418}\ } (\bibinfo {year} {2023})}\BibitemShut {NoStop}%
\bibitem [{\citenamefont {Guo}\ \emph {et~al.}(2023{\natexlab{b}})\citenamefont {Guo}, \citenamefont {Gu}, \citenamefont {Gao},\ and\ \citenamefont {Lu}}]{zhongyi-Lu-altermagnetic-FE}%
  \BibitemOpen
  \bibfield  {author} {\bibinfo {author} {\bibfnamefont {P.-J.}\ \bibnamefont {Guo}}, \bibinfo {author} {\bibfnamefont {Y.}~\bibnamefont {Gu}}, \bibinfo {author} {\bibfnamefont {Z.-F.}\ \bibnamefont {Gao}},\ and\ \bibinfo {author} {\bibfnamefont {Z.-Y.}\ \bibnamefont {Lu}},\ }\bibfield  {title} {\bibinfo {title} {Altermagnetic ferroelectric LiFe$_2$F$_6$ and spin-triplet excitonic insulator phase},\ }\href@noop {} {\bibfield  {journal} {\bibinfo  {journal} {arXiv preprint arXiv:2312.13911}\ } (\bibinfo {year} {2023}{\natexlab{b}})}\BibitemShut {NoStop}%
\bibitem [{\citenamefont {Qu}\ \emph {et~al.}(2024)\citenamefont {Qu}, \citenamefont {Gao}, \citenamefont {Sun}, \citenamefont {Liu}, \citenamefont {Guo},\ and\ \citenamefont {Lu}}]{Zhongyi-Lu-NiF3}%
  \BibitemOpen
  \bibfield  {author} {\bibinfo {author} {\bibfnamefont {S.}~\bibnamefont {Qu}}, \bibinfo {author} {\bibfnamefont {Z.-F.}\ \bibnamefont {Gao}}, \bibinfo {author} {\bibfnamefont {H.}~\bibnamefont {Sun}}, \bibinfo {author} {\bibfnamefont {K.}~\bibnamefont {Liu}}, \bibinfo {author} {\bibfnamefont {P.-J.}\ \bibnamefont {Guo}},\ and\ \bibinfo {author} {\bibfnamefont {Z.-Y.}\ \bibnamefont {Lu}},\ }\bibfield  {title} {\bibinfo {title} {Extremely strong spin-orbit coupling effect in light element altermagnetic materials},\ }\href@noop {} {\bibfield  {journal} {\bibinfo  {journal} {arXiv preprint arXiv:2401.11065}\ } (\bibinfo {year} {2024})}\BibitemShut {NoStop}%
\bibitem [{\citenamefont {Hu}\ \emph {et~al.}(2024)\citenamefont {Hu}, \citenamefont {Cheng}, \citenamefont {Huang},\ and\ \citenamefont {Liu}}]{liu2024}%
  \BibitemOpen
  \bibfield  {author} {\bibinfo {author} {\bibfnamefont {M.}~\bibnamefont {Hu}}, \bibinfo {author} {\bibfnamefont {X.}~\bibnamefont {Cheng}}, \bibinfo {author} {\bibfnamefont {Z.}~\bibnamefont {Huang}},\ and\ \bibinfo {author} {\bibfnamefont {J.}~\bibnamefont {Liu}},\ }\bibfield  {title} {\bibinfo {title} {Catalogue of $ c $-paired spin-valley locking in antiferromagnetic systems},\ }\href@noop {} {\bibfield  {journal} {\bibinfo  {journal} {arXiv preprint arXiv:2407.02319}\ } (\bibinfo {year} {2024})}\BibitemShut {NoStop}%
\bibitem [{\citenamefont {Krempask{\`y}}\ \emph {et~al.}(2024)\citenamefont {Krempask{\`y}}, \citenamefont {{\v{S}}mejkal}, \citenamefont {D’souza}, \citenamefont {Hajlaoui}, \citenamefont {Springholz}, \citenamefont {Uhl{\'\i}{\v{r}}ov{\'a}}, \citenamefont {Alarab}, \citenamefont {Constantinou}, \citenamefont {Strocov}, \citenamefont {Usanov} \emph {et~al.}}]{MnTe-ARPES1-Nature}%
  \BibitemOpen
  \bibfield  {author} {\bibinfo {author} {\bibfnamefont {J.}~\bibnamefont {Krempask{\`y}}}, \bibinfo {author} {\bibfnamefont {L.}~\bibnamefont {{\v{S}}mejkal}}, \bibinfo {author} {\bibfnamefont {S.}~\bibnamefont {D’souza}}, \bibinfo {author} {\bibfnamefont {M.}~\bibnamefont {Hajlaoui}}, \bibinfo {author} {\bibfnamefont {G.}~\bibnamefont {Springholz}}, \bibinfo {author} {\bibfnamefont {K.}~\bibnamefont {Uhl{\'\i}{\v{r}}ov{\'a}}}, \bibinfo {author} {\bibfnamefont {F.}~\bibnamefont {Alarab}}, \bibinfo {author} {\bibfnamefont {P.}~\bibnamefont {Constantinou}}, \bibinfo {author} {\bibfnamefont {V.}~\bibnamefont {Strocov}}, \bibinfo {author} {\bibfnamefont {D.}~\bibnamefont {Usanov}}, \emph {et~al.},\ }\bibfield  {title} {\bibinfo {title} {Altermagnetic lifting of kramers spin degeneracy},\ }\href@noop {} {\bibfield  {journal} {\bibinfo  {journal} {Nature}\ }\textbf {\bibinfo {volume} {626}},\ \bibinfo {pages} {517} (\bibinfo {year} {2024})}\BibitemShut {NoStop}%
\bibitem [{\citenamefont {Gonzalez~Betancourt}\ \emph {et~al.}(2023{\natexlab{a}})\citenamefont {Gonzalez~Betancourt}, \citenamefont {Zub{\'a}{\v{c}}}, \citenamefont {Gonzalez-Hernandez}, \citenamefont {Geishendorf}, \citenamefont {{\v{S}}ob{\'a}{\v{n}}}, \citenamefont {Springholz}, \citenamefont {Olejn{\'\i}k}, \citenamefont {{\v{S}}mejkal}, \citenamefont {Sinova}, \citenamefont {Jungwirth} \emph {et~al.}}]{MnTe-ARPES2-PRL}%
  \BibitemOpen
  \bibfield  {author} {\bibinfo {author} {\bibfnamefont {R.}~\bibnamefont {Gonzalez~Betancourt}}, \bibinfo {author} {\bibfnamefont {J.}~\bibnamefont {Zub{\'a}{\v{c}}}}, \bibinfo {author} {\bibfnamefont {R.}~\bibnamefont {Gonzalez-Hernandez}}, \bibinfo {author} {\bibfnamefont {K.}~\bibnamefont {Geishendorf}}, \bibinfo {author} {\bibfnamefont {Z.}~\bibnamefont {{\v{S}}ob{\'a}{\v{n}}}}, \bibinfo {author} {\bibfnamefont {G.}~\bibnamefont {Springholz}}, \bibinfo {author} {\bibfnamefont {K.}~\bibnamefont {Olejn{\'\i}k}}, \bibinfo {author} {\bibfnamefont {L.}~\bibnamefont {{\v{S}}mejkal}}, \bibinfo {author} {\bibfnamefont {J.}~\bibnamefont {Sinova}}, \bibinfo {author} {\bibfnamefont {T.}~\bibnamefont {Jungwirth}}, \emph {et~al.},\ }\bibfield  {title} {\bibinfo {title} {Spontaneous anomalous Hall effect arising from an unconventional compensated magnetic phase in a semiconductor},\ }\href@noop {} {\bibfield  {journal} {\bibinfo  {journal} {Physical Review Letters}\ }\textbf {\bibinfo {volume} {130}},\ \bibinfo
  {pages} {036702} (\bibinfo {year} {2023}{\natexlab{a}})}\BibitemShut {NoStop}%
\bibitem [{\citenamefont {Osumi}\ \emph {et~al.}(2024)\citenamefont {Osumi}, \citenamefont {Souma}, \citenamefont {Aoyama}, \citenamefont {Yamauchi}, \citenamefont {Honma}, \citenamefont {Nakayama}, \citenamefont {Takahashi}, \citenamefont {Ohgushi},\ and\ \citenamefont {Sato}}]{MnTe-ARPES3-PRb}%
  \BibitemOpen
  \bibfield  {author} {\bibinfo {author} {\bibfnamefont {T.}~\bibnamefont {Osumi}}, \bibinfo {author} {\bibfnamefont {S.}~\bibnamefont {Souma}}, \bibinfo {author} {\bibfnamefont {T.}~\bibnamefont {Aoyama}}, \bibinfo {author} {\bibfnamefont {K.}~\bibnamefont {Yamauchi}}, \bibinfo {author} {\bibfnamefont {A.}~\bibnamefont {Honma}}, \bibinfo {author} {\bibfnamefont {K.}~\bibnamefont {Nakayama}}, \bibinfo {author} {\bibfnamefont {T.}~\bibnamefont {Takahashi}}, \bibinfo {author} {\bibfnamefont {K.}~\bibnamefont {Ohgushi}},\ and\ \bibinfo {author} {\bibfnamefont {T.}~\bibnamefont {Sato}},\ }\bibfield  {title} {\bibinfo {title} {Observation of a giant band splitting in altermagnetic MnTe},\ }\href@noop {} {\bibfield  {journal} {\bibinfo  {journal} {Physical Review B}\ }\textbf {\bibinfo {volume} {109}},\ \bibinfo {pages} {115102} (\bibinfo {year} {2024})}\BibitemShut {NoStop}%
 \bibitem [{\citenamefont {Nagaosa}\ \emph {et~al.}(2010)\citenamefont {Nagaosa}, \citenamefont {Sinova}, \citenamefont {Onoda}, \citenamefont {MacDonald},\ and\ \citenamefont {Ong}}]{AHC2}%
  \BibitemOpen
  \bibfield  {author} {\bibinfo {author} {\bibfnamefont {N.}~\bibnamefont {Nagaosa}}, \bibinfo {author} {\bibfnamefont {J.}~\bibnamefont {Sinova}}, \bibinfo {author} {\bibfnamefont {S.}~\bibnamefont {Onoda}}, \bibinfo {author} {\bibfnamefont {A.~H.}\ \bibnamefont {MacDonald}},\ and\ \bibinfo {author} {\bibfnamefont {N.~P.}\ \bibnamefont {Ong}},\ }\bibfield  {title} {\bibinfo {title} {Anomalous Hall effect},\ }\href {https://doi.org/10.1103/RevModPhys.82.1539} {\bibfield  {journal} {\bibinfo  {journal} {Rev. Mod. Phys.}\ }\textbf {\bibinfo {volume} {82}},\ \bibinfo {pages} {1539} (\bibinfo {year} {2010})}\BibitemShut {NoStop}%
\bibitem [{\citenamefont {Xiao}\ \emph {et~al.}(2010)\citenamefont {Xiao}, \citenamefont {Chang},\ and\ \citenamefont {Niu}}]{Xiao-Di-RMP}%
  \BibitemOpen
  \bibfield  {author} {\bibinfo {author} {\bibfnamefont {D.}~\bibnamefont {Xiao}}, \bibinfo {author} {\bibfnamefont {M.-C.}\ \bibnamefont {Chang}},\ and\ \bibinfo {author} {\bibfnamefont {Q.}~\bibnamefont {Niu}},\ }\bibfield  {title} {\bibinfo {title} {Berry phase effects on electronic properties},\ }\href {https://doi.org/10.1103/RevModPhys.82.1959} {\bibfield  {journal} {\bibinfo  {journal} {Rev. Mod. Phys.}\ }\textbf {\bibinfo {volume} {82}},\ \bibinfo {pages} {1959} (\bibinfo {year} {2010})}\BibitemShut {NoStop}%
\bibitem [{\citenamefont {Zhang}\ \emph {et~al.}(2014)\citenamefont {Zhang}, \citenamefont {Liu}, \citenamefont {Luo}, \citenamefont {Freeman},\ and\ \citenamefont {Zunger}}]{BC1}%
  \BibitemOpen
  \bibfield  {author} {\bibinfo {author} {\bibfnamefont {X.}~\bibnamefont {Zhang}}, \bibinfo {author} {\bibfnamefont {Q.}~\bibnamefont {Liu}}, \bibinfo {author} {\bibfnamefont {J.-W.}\ \bibnamefont {Luo}}, \bibinfo {author} {\bibfnamefont {A.~J.}\ \bibnamefont {Freeman}},\ and\ \bibinfo {author} {\bibfnamefont {A.}~\bibnamefont {Zunger}},\ }\bibfield  {title} {\bibinfo {title} {Hidden spin polarization in inversion-symmetric bulk crystals},\ }\href@noop {} {\bibfield  {journal} {\bibinfo  {journal} {Nature Physics}\ }\textbf {\bibinfo {volume} {10}},\ \bibinfo {pages} {387} (\bibinfo {year} {2014})}\BibitemShut {NoStop}%
\bibitem [{\citenamefont {{\v{S}}mejkal}\ \emph {et~al.}(2017)\citenamefont {{\v{S}}mejkal}, \citenamefont {{\v{Z}}elezn{\`y}}, \citenamefont {Sinova},\ and\ \citenamefont {Jungwirth}}]{BC2}%
  \BibitemOpen
  \bibfield  {author} {\bibinfo {author} {\bibfnamefont {L.}~\bibnamefont {{\v{S}}mejkal}}, \bibinfo {author} {\bibfnamefont {J.}~\bibnamefont {{\v{Z}}elezn{\`y}}}, \bibinfo {author} {\bibfnamefont {J.}~\bibnamefont {Sinova}},\ and\ \bibinfo {author} {\bibfnamefont {T.}~\bibnamefont {Jungwirth}},\ }\bibfield  {title} {\bibinfo {title} {Electric control of dirac quasiparticles by spin-orbit torque in an antiferromagnet},\ }\href@noop {} {\bibfield  {journal} {\bibinfo  {journal} {Physical review letters}\ }\textbf {\bibinfo {volume} {118}},\ \bibinfo {pages} {106402} (\bibinfo {year} {2017})}\BibitemShut {NoStop}%
\bibitem [{\citenamefont {Feng}\ \emph {et~al.}(2022)\citenamefont {Feng}, \citenamefont {Zhou}, \citenamefont {{\v{S}}mejkal}, \citenamefont {Wu}, \citenamefont {Zhu}, \citenamefont {Guo}, \citenamefont {Gonz{\'a}lez-Hern{\'a}ndez}, \citenamefont {Wang}, \citenamefont {Yan}, \citenamefont {Qin} \emph {et~al.}}]{AHC_EXP_RuO2}%
  \BibitemOpen
  \bibfield  {author} {\bibinfo {author} {\bibfnamefont {Z.}~\bibnamefont {Feng}}, \bibinfo {author} {\bibfnamefont {X.}~\bibnamefont {Zhou}}, \bibinfo {author} {\bibfnamefont {L.}~\bibnamefont {{\v{S}}mejkal}}, \bibinfo {author} {\bibfnamefont {L.}~\bibnamefont {Wu}}, \bibinfo {author} {\bibfnamefont {Z.}~\bibnamefont {Zhu}}, \bibinfo {author} {\bibfnamefont {H.}~\bibnamefont {Guo}}, \bibinfo {author} {\bibfnamefont {R.}~\bibnamefont {Gonz{\'a}lez-Hern{\'a}ndez}}, \bibinfo {author} {\bibfnamefont {X.}~\bibnamefont {Wang}}, \bibinfo {author} {\bibfnamefont {H.}~\bibnamefont {Yan}}, \bibinfo {author} {\bibfnamefont {P.}~\bibnamefont {Qin}}, \emph {et~al.},\ }\bibfield  {title} {\bibinfo {title} {An anomalous Hall effect in altermagnetic ruthenium dioxide},\ }\href@noop {} {\bibfield  {journal} {\bibinfo  {journal} {Nature Electronics}\ }\textbf {\bibinfo {volume} {5}},\ \bibinfo {pages} {735} (\bibinfo {year} {2022})}\BibitemShut {NoStop}%
\bibitem [{\citenamefont {Reichlova}\ \emph {et~al.}(2024)\citenamefont {Reichlova}, \citenamefont {Lopes~Seeger}, \citenamefont {Gonz{\'a}lez-Hern{\'a}ndez}, \citenamefont {Kounta}, \citenamefont {Schlitz}, \citenamefont {Kriegner}, \citenamefont {Ritzinger}, \citenamefont {Lammel}, \citenamefont {Leivisk{\"a}}, \citenamefont {Birk~Hellenes} \emph {et~al.}}]{nc_Mn5Si3}%
  \BibitemOpen
  \bibfield  {author} {\bibinfo {author} {\bibfnamefont {H.}~\bibnamefont {Reichlova}}, \bibinfo {author} {\bibfnamefont {R.}~\bibnamefont {Lopes~Seeger}}, \bibinfo {author} {\bibfnamefont {R.}~\bibnamefont {Gonz{\'a}lez-Hern{\'a}ndez}}, \bibinfo {author} {\bibfnamefont {I.}~\bibnamefont {Kounta}}, \bibinfo {author} {\bibfnamefont {R.}~\bibnamefont {Schlitz}}, \bibinfo {author} {\bibfnamefont {D.}~\bibnamefont {Kriegner}}, \bibinfo {author} {\bibfnamefont {P.}~\bibnamefont {Ritzinger}}, \bibinfo {author} {\bibfnamefont {M.}~\bibnamefont {Lammel}}, \bibinfo {author} {\bibfnamefont {M.}~\bibnamefont {Leivisk{\"a}}}, \bibinfo {author} {\bibfnamefont {A.}~\bibnamefont {Birk~Hellenes}}, \emph {et~al.},\ }\bibfield  {title} {\bibinfo {title} {Observation of a spontaneous anomalous hall response in the Mn$_5$Si$_3$ d-wave altermagnet candidate},\ }\href@noop {} {\bibfield  {journal} {\bibinfo  {journal} {Nature Communications}\ }\textbf {\bibinfo {volume} {15}},\ \bibinfo {pages} {4961} (\bibinfo {year}
  {2024})}\BibitemShut {NoStop}%
\bibitem [{\citenamefont {Gonzalez~Betancourt}\ \emph {et~al.}(2023{\natexlab{b}})\citenamefont {Gonzalez~Betancourt}, \citenamefont {Zub{\'a}{\v{c}}}, \citenamefont {Gonzalez-Hernandez}, \citenamefont {Geishendorf}, \citenamefont {{\v{S}}ob{\'a}{\v{n}}}, \citenamefont {Springholz}, \citenamefont {Olejn{\'\i}k}, \citenamefont {{\v{S}}mejkal}, \citenamefont {Sinova}, \citenamefont {Jungwirth} \emph {et~al.}}]{AHC_EXP_MnTe}%
  \BibitemOpen
  \bibfield  {author} {\bibinfo {author} {\bibfnamefont {R.}~\bibnamefont {Gonzalez~Betancourt}}, \bibinfo {author} {\bibfnamefont {J.}~\bibnamefont {Zub{\'a}{\v{c}}}}, \bibinfo {author} {\bibfnamefont {R.}~\bibnamefont {Gonzalez-Hernandez}}, \bibinfo {author} {\bibfnamefont {K.}~\bibnamefont {Geishendorf}}, \bibinfo {author} {\bibfnamefont {Z.}~\bibnamefont {{\v{S}}ob{\'a}{\v{n}}}}, \bibinfo {author} {\bibfnamefont {G.}~\bibnamefont {Springholz}}, \bibinfo {author} {\bibfnamefont {K.}~\bibnamefont {Olejn{\'\i}k}}, \bibinfo {author} {\bibfnamefont {L.}~\bibnamefont {{\v{S}}mejkal}}, \bibinfo {author} {\bibfnamefont {J.}~\bibnamefont {Sinova}}, \bibinfo {author} {\bibfnamefont {T.}~\bibnamefont {Jungwirth}}, \emph {et~al.},\ }\bibfield  {title} {\bibinfo {title} {Spontaneous anomalous Hall effect arising from an unconventional compensated magnetic phase in a semiconductor},\ }\href@noop {} {\bibfield  {journal} {\bibinfo  {journal} {Physical Review Letters}\ }\textbf {\bibinfo {volume} {130}},\ \bibinfo
  {pages} {036702} (\bibinfo {year} {2023}{\natexlab{b}})}\BibitemShut {NoStop}%
\bibitem [{\citenamefont {Fedchenko}\ \emph {et~al.}(2024)\citenamefont {Fedchenko}, \citenamefont {Min{\'a}r}, \citenamefont {Akashdeep}, \citenamefont {D’Souza}, \citenamefont {Vasilyev}, \citenamefont {Tkach}, \citenamefont {Odenbreit}, \citenamefont {Nguyen}, \citenamefont {Kutnyakhov}, \citenamefont {Wind} \emph {et~al.}}]{RuO2-AM1}%
  \BibitemOpen
  \bibfield  {author} {\bibinfo {author} {\bibfnamefont {O.}~\bibnamefont {Fedchenko}}, \bibinfo {author} {\bibfnamefont {J.}~\bibnamefont {Min{\'a}r}}, \bibinfo {author} {\bibfnamefont {A.}~\bibnamefont {Akashdeep}}, \bibinfo {author} {\bibfnamefont {S.~W.}\ \bibnamefont {D’Souza}}, \bibinfo {author} {\bibfnamefont {D.}~\bibnamefont {Vasilyev}}, \bibinfo {author} {\bibfnamefont {O.}~\bibnamefont {Tkach}}, \bibinfo {author} {\bibfnamefont {L.}~\bibnamefont {Odenbreit}}, \bibinfo {author} {\bibfnamefont {Q.}~\bibnamefont {Nguyen}}, \bibinfo {author} {\bibfnamefont {D.}~\bibnamefont {Kutnyakhov}}, \bibinfo {author} {\bibfnamefont {N.}~\bibnamefont {Wind}}, \emph {et~al.},\ }\bibfield  {title} {\bibinfo {title} {Observation of time-reversal symmetry breaking in the band structure of altermagnetic RuO$_2$},\ }\href@noop {} {\bibfield  {journal} {\bibinfo  {journal} {Science advances}\ }\textbf {\bibinfo {volume} {10}},\ \bibinfo {pages} {eadj4883} (\bibinfo {year} {2024})}\BibitemShut {NoStop}%
\bibitem [{\citenamefont {Zhu}\ \emph {et~al.}(2019)\citenamefont {Zhu}, \citenamefont {Strempfer}, \citenamefont {Rao}, \citenamefont {Occhialini}, \citenamefont {Pelliciari}, \citenamefont {Choi}, \citenamefont {Kawaguchi}, \citenamefont {You}, \citenamefont {Mitchell}, \citenamefont {Shao-Horn} \emph {et~al.}}]{RuO2-AM2}%
  \BibitemOpen
  \bibfield  {author} {\bibinfo {author} {\bibfnamefont {Z.}~\bibnamefont {Zhu}}, \bibinfo {author} {\bibfnamefont {J.}~\bibnamefont {Strempfer}}, \bibinfo {author} {\bibfnamefont {R.}~\bibnamefont {Rao}}, \bibinfo {author} {\bibfnamefont {C.}~\bibnamefont {Occhialini}}, \bibinfo {author} {\bibfnamefont {J.}~\bibnamefont {Pelliciari}}, \bibinfo {author} {\bibfnamefont {Y.}~\bibnamefont {Choi}}, \bibinfo {author} {\bibfnamefont {T.}~\bibnamefont {Kawaguchi}}, \bibinfo {author} {\bibfnamefont {H.}~\bibnamefont {You}}, \bibinfo {author} {\bibfnamefont {J.}~\bibnamefont {Mitchell}}, \bibinfo {author} {\bibfnamefont {Y.}~\bibnamefont {Shao-Horn}}, \emph {et~al.},\ }\bibfield  {title} {\bibinfo {title} {Anomalous antiferromagnetism in metallic RuO$_2$ determined by resonant x-ray scattering},\ }\href@noop {} {\bibfield  {journal} {\bibinfo  {journal} {Physical review letters}\ }\textbf {\bibinfo {volume} {122}},\ \bibinfo {pages} {017202} (\bibinfo {year} {2019})}\BibitemShut {NoStop}%
\bibitem [{\citenamefont {Lin}\ \emph {et~al.}(2024)\citenamefont {Lin}, \citenamefont {Chen}, \citenamefont {Lu}, \citenamefont {Liang}, \citenamefont {Feng}, \citenamefont {Yamagami}, \citenamefont {Osiecki}, \citenamefont {Leandersson}, \citenamefont {Thiagarajan}, \citenamefont {Liu} \emph {et~al.}}]{RuO2-AM3}%
  \BibitemOpen
  \bibfield  {author} {\bibinfo {author} {\bibfnamefont {Z.}~\bibnamefont {Lin}}, \bibinfo {author} {\bibfnamefont {D.}~\bibnamefont {Chen}}, \bibinfo {author} {\bibfnamefont {W.}~\bibnamefont {Lu}}, \bibinfo {author} {\bibfnamefont {X.}~\bibnamefont {Liang}}, \bibinfo {author} {\bibfnamefont {S.}~\bibnamefont {Feng}}, \bibinfo {author} {\bibfnamefont {K.}~\bibnamefont {Yamagami}}, \bibinfo {author} {\bibfnamefont {J.}~\bibnamefont {Osiecki}}, \bibinfo {author} {\bibfnamefont {M.}~\bibnamefont {Leandersson}}, \bibinfo {author} {\bibfnamefont {B.}~\bibnamefont {Thiagarajan}}, \bibinfo {author} {\bibfnamefont {J.}~\bibnamefont {Liu}}, \emph {et~al.},\ }\bibfield  {title} {\bibinfo {title} {Observation of giant spin splitting and d-wave spin texture in room temperature altermagnet RuO$_2$},\ }\href@noop {} {\bibfield  {journal} {\bibinfo  {journal} {arXiv preprint arXiv:2402.04995}\ } (\bibinfo {year} {2024})}\BibitemShut {NoStop}%
\bibitem [{\citenamefont {Huang}\ \emph {et~al.}(2024)\citenamefont {Huang}, \citenamefont {Lai}, \citenamefont {Zhan}, \citenamefont {Yu}, \citenamefont {Chen}, \citenamefont {Liu}, \citenamefont {Chen},\ and\ \citenamefont {Sun}}]{RuO2-huang}%
  \BibitemOpen
  \bibfield  {author} {\bibinfo {author} {\bibfnamefont {Y.}~\bibnamefont {Huang}}, \bibinfo {author} {\bibfnamefont {J.}~\bibnamefont {Lai}}, \bibinfo {author} {\bibfnamefont {J.}~\bibnamefont {Zhan}}, \bibinfo {author} {\bibfnamefont {T.}~\bibnamefont {Yu}}, \bibinfo {author} {\bibfnamefont {R.}~\bibnamefont {Chen}}, \bibinfo {author} {\bibfnamefont {P.}~\bibnamefont {Liu}}, \bibinfo {author} {\bibfnamefont {X.-Q.}\ \bibnamefont {Chen}},\ and\ \bibinfo {author} {\bibfnamefont {Y.}~\bibnamefont {Sun}},\ }\bibfield  {title} {\bibinfo {title} {Ab initio study of quantum oscillations in altermagnetic and nonmagnetic phases of RuO$_2$},\ }\href@noop {} {\bibfield  {journal} {\bibinfo  {journal} {Physical Review B}\ }\textbf {\bibinfo {volume} {110}},\ \bibinfo {pages} {144410} (\bibinfo {year} {2024})}\BibitemShut {NoStop}%
\bibitem [{\citenamefont {Hiraishi}\ \emph {et~al.}(2024)\citenamefont {Hiraishi}, \citenamefont {Okabe}, \citenamefont {Koda}, \citenamefont {Kadono}, \citenamefont {Muroi}, \citenamefont {Hirai},\ and\ \citenamefont {Hiroi}}]{RuO2-nonmagnetic1}%
  \BibitemOpen
  \bibfield  {author} {\bibinfo {author} {\bibfnamefont {M.}~\bibnamefont {Hiraishi}}, \bibinfo {author} {\bibfnamefont {H.}~\bibnamefont {Okabe}}, \bibinfo {author} {\bibfnamefont {A.}~\bibnamefont {Koda}}, \bibinfo {author} {\bibfnamefont {R.}~\bibnamefont {Kadono}}, \bibinfo {author} {\bibfnamefont {T.}~\bibnamefont {Muroi}}, \bibinfo {author} {\bibfnamefont {D.}~\bibnamefont {Hirai}},\ and\ \bibinfo {author} {\bibfnamefont {Z.}~\bibnamefont {Hiroi}},\ }\bibfield  {title} {\bibinfo {title} {Nonmagnetic ground state in RuO$_2$ revealed by muon spin rotation},\ }\href@noop {} {\bibfield  {journal} {\bibinfo  {journal} {Physical Review Letters}\ }\textbf {\bibinfo {volume} {132}},\ \bibinfo {pages} {166702} (\bibinfo {year} {2024})}\BibitemShut {NoStop}%
\bibitem [{\citenamefont {Ke{\ss}ler}\ \emph {et~al.}(2024)\citenamefont {Ke{\ss}ler}, \citenamefont {Garcia-Gassull}, \citenamefont {Suter}, \citenamefont {Prokscha}, \citenamefont {Salman}, \citenamefont {Khalyavin}, \citenamefont {Manuel}, \citenamefont {Orlandi}, \citenamefont {Mazin}, \citenamefont {Valent{\'\i}},\ and\ \citenamefont {Moser}}]{RuO2-nonmagnetic2}%
  \BibitemOpen
  \bibfield  {author} {\bibinfo {author} {\bibfnamefont {P.}~\bibnamefont {Ke{\ss}ler}}, \bibinfo {author} {\bibfnamefont {L.}~\bibnamefont {Garcia-Gassull}}, \bibinfo {author} {\bibfnamefont {A.}~\bibnamefont {Suter}}, \bibinfo {author} {\bibfnamefont {T.}~\bibnamefont {Prokscha}}, \bibinfo {author} {\bibfnamefont {Z.}~\bibnamefont {Salman}}, \bibinfo {author} {\bibfnamefont {D.}~\bibnamefont {Khalyavin}}, \bibinfo {author} {\bibfnamefont {P.}~\bibnamefont {Manuel}}, \bibinfo {author} {\bibfnamefont {F.}~\bibnamefont {Orlandi}}, \bibinfo {author} {\bibfnamefont {I.~I.}\ \bibnamefont {Mazin}}, \bibinfo {author} {\bibfnamefont {R.}~\bibnamefont {Valent{\'\i}}},\ and\ \bibinfo {author} {\bibfnamefont {S.}~\bibnamefont {Moser}},\ }\bibfield  {title} {\bibinfo {title} {Absence of magnetic order in RuO$_2$: insights from $\mu$ sr spectroscopy and neutron diffraction},\ }\href@noop {} {\bibfield  {journal} {\bibinfo  {journal} {npj Spintronics}\ }\textbf {\bibinfo {volume} {2}},\ \bibinfo {pages} {50} (\bibinfo {year}
  {2024})}\BibitemShut {NoStop}%
\bibitem [{\citenamefont {Liu}\ \emph {et~al.}(2024)\citenamefont {Liu}, \citenamefont {Zhan}, \citenamefont {Li}, \citenamefont {Liu}, \citenamefont {Cheng}, \citenamefont {Shi}, \citenamefont {Deng}, \citenamefont {Zhang}, \citenamefont {Li}, \citenamefont {Ding}, \citenamefont {Jiang}, \citenamefont {Ye}, \citenamefont {Liu}, \citenamefont {Jiang}, \citenamefont {Wang}, \citenamefont {Li}, \citenamefont {Xie}, \citenamefont {Wang}, \citenamefont {Qiao}, \citenamefont {Wen}, \citenamefont {Sun},\ and\ \citenamefont {Shen}}]{RuO2-nonmagnetic3}%
  \BibitemOpen
  \bibfield  {author} {\bibinfo {author} {\bibfnamefont {J.}~\bibnamefont {Liu}}, \bibinfo {author} {\bibfnamefont {J.}~\bibnamefont {Zhan}}, \bibinfo {author} {\bibfnamefont {T.}~\bibnamefont {Li}}, \bibinfo {author} {\bibfnamefont {J.}~\bibnamefont {Liu}}, \bibinfo {author} {\bibfnamefont {S.}~\bibnamefont {Cheng}}, \bibinfo {author} {\bibfnamefont {Y.}~\bibnamefont {Shi}}, \bibinfo {author} {\bibfnamefont {L.}~\bibnamefont {Deng}}, \bibinfo {author} {\bibfnamefont {M.}~\bibnamefont {Zhang}}, \bibinfo {author} {\bibfnamefont {C.}~\bibnamefont {Li}}, \bibinfo {author} {\bibfnamefont {J.}~\bibnamefont {Ding}}, \bibinfo {author} {\bibfnamefont {Q.}~\bibnamefont {Jiang}}, \bibinfo {author} {\bibfnamefont {M.}~\bibnamefont {Ye}}, \bibinfo {author} {\bibfnamefont {Z.}~\bibnamefont {Liu}}, \bibinfo {author} {\bibfnamefont {Z.}~\bibnamefont {Jiang}}, \bibinfo {author} {\bibfnamefont {S.}~\bibnamefont {Wang}}, \bibinfo {author} {\bibfnamefont {Q.}~\bibnamefont {Li}}, \bibinfo {author} {\bibfnamefont
  {Y.}~\bibnamefont {Xie}}, \bibinfo {author} {\bibfnamefont {Y.}~\bibnamefont {Wang}}, \bibinfo {author} {\bibfnamefont {S.}~\bibnamefont {Qiao}}, \bibinfo {author} {\bibfnamefont {J.}~\bibnamefont {Wen}}, \bibinfo {author} {\bibfnamefont {Y.}~\bibnamefont {Sun}},\ and\ \bibinfo {author} {\bibfnamefont {D.}~\bibnamefont {Shen}},\ }\bibfield  {title} {\bibinfo {title} {Absence of altermagnetic spin splitting character in rutile oxide ${\mathrm{RuO}}_{2}$},\ }\href {https://doi.org/10.1103/PhysRevLett.133.176401} {\bibfield  {journal} {\bibinfo  {journal} {Phys. Rev. Lett.}\ }\textbf {\bibinfo {volume} {133}},\ \bibinfo {pages} {176401} (\bibinfo {year} {2024})}\BibitemShut {NoStop}%
\bibitem [{\citenamefont {Allen}\ \emph {et~al.}(1977)\citenamefont {Allen}, \citenamefont {Lucovsky},\ and\ \citenamefont {Mikkelsen~Jr}}]{MnTe-band_gap1}%
  \BibitemOpen
  \bibfield  {author} {\bibinfo {author} {\bibfnamefont {J.}~\bibnamefont {Allen}}, \bibinfo {author} {\bibfnamefont {G.}~\bibnamefont {Lucovsky}},\ and\ \bibinfo {author} {\bibfnamefont {J.}~\bibnamefont {Mikkelsen~Jr}},\ }\bibfield  {title} {\bibinfo {title} {Optical properties and electronic structure of crossroads material MnTe},\ }\href@noop {} {\bibfield  {journal} {\bibinfo  {journal} {Solid State Communications}\ }\textbf {\bibinfo {volume} {24}},\ \bibinfo {pages} {367} (\bibinfo {year} {1977})}\BibitemShut {NoStop}%
\bibitem [{\citenamefont {Ferrer-Roca}\ \emph {et~al.}(2000)\citenamefont {Ferrer-Roca}, \citenamefont {Segura}, \citenamefont {Reig},\ and\ \citenamefont {Mu\~noz}}]{MnTe-band_gap2}%
  \BibitemOpen
  \bibfield  {author} {\bibinfo {author} {\bibfnamefont {C.}~\bibnamefont {Ferrer-Roca}}, \bibinfo {author} {\bibfnamefont {A.}~\bibnamefont {Segura}}, \bibinfo {author} {\bibfnamefont {C.}~\bibnamefont {Reig}},\ and\ \bibinfo {author} {\bibfnamefont {V.}~\bibnamefont {Mu\~noz}},\ }\bibfield  {title} {\bibinfo {title} {Temperature and pressure dependence of the optical absorption in hexagonal MnTe},\ }\href {https://doi.org/10.1103/PhysRevB.61.13679} {\bibfield  {journal} {\bibinfo  {journal} {Phys. Rev. B}\ }\textbf {\bibinfo {volume} {61}},\ \bibinfo {pages} {13679} (\bibinfo {year} {2000})}\BibitemShut {NoStop}%
\bibitem [{\citenamefont {S{\"u}rgers}\ \emph {et~al.}(2024)\citenamefont {S{\"u}rgers}, \citenamefont {Fischer}, \citenamefont {Campos}, \citenamefont {Hellenes}, \citenamefont {{\v{S}}mejkal}, \citenamefont {Sinova}, \citenamefont {Merz}, \citenamefont {Wolf},\ and\ \citenamefont {Wernsdorfer}}]{Mn5Si3-magnetic_state1}%
  \BibitemOpen
  \bibfield  {author} {\bibinfo {author} {\bibfnamefont {C.}~\bibnamefont {S{\"u}rgers}}, \bibinfo {author} {\bibfnamefont {G.}~\bibnamefont {Fischer}}, \bibinfo {author} {\bibfnamefont {W.~H.}\ \bibnamefont {Campos}}, \bibinfo {author} {\bibfnamefont {A.~B.}\ \bibnamefont {Hellenes}}, \bibinfo {author} {\bibfnamefont {L.}~\bibnamefont {{\v{S}}mejkal}}, \bibinfo {author} {\bibfnamefont {J.}~\bibnamefont {Sinova}}, \bibinfo {author} {\bibfnamefont {M.}~\bibnamefont {Merz}}, \bibinfo {author} {\bibfnamefont {T.}~\bibnamefont {Wolf}},\ and\ \bibinfo {author} {\bibfnamefont {W.}~\bibnamefont {Wernsdorfer}},\ }\bibfield  {title} {\bibinfo {title} {Anomalous Nernst effect in the noncollinear antiferromagnet Mn$_5$Si$_3$},\ }\href@noop {} {\bibfield  {journal} {\bibinfo  {journal} {Communications Materials}\ }\textbf {\bibinfo {volume} {5}},\ \bibinfo {pages} {176} (\bibinfo {year} {2024})}\BibitemShut {NoStop}%
\bibitem [{\citenamefont {Gottschilch}\ \emph {et~al.}(2012)\citenamefont {Gottschilch}, \citenamefont {Gourdon}, \citenamefont {Persson}, \citenamefont {de~la Cruz}, \citenamefont {Petricek},\ and\ \citenamefont {Brueckel}}]{Mn5Si3-magnetic_state2}%
  \BibitemOpen
  \bibfield  {author} {\bibinfo {author} {\bibfnamefont {M.}~\bibnamefont {Gottschilch}}, \bibinfo {author} {\bibfnamefont {O.}~\bibnamefont {Gourdon}}, \bibinfo {author} {\bibfnamefont {J.}~\bibnamefont {Persson}}, \bibinfo {author} {\bibfnamefont {C.}~\bibnamefont {de~la Cruz}}, \bibinfo {author} {\bibfnamefont {V.}~\bibnamefont {Petricek}},\ and\ \bibinfo {author} {\bibfnamefont {T.}~\bibnamefont {Brueckel}},\ }\bibfield  {title} {\bibinfo {title} {Study of the antiferromagnetism of Mn$_5$Si$_3$: an inverse magnetocaloric effect material},\ }\href@noop {} {\bibfield  {journal} {\bibinfo  {journal} {Journal of materials chemistry}\ }\textbf {\bibinfo {volume} {22}},\ \bibinfo {pages} {15275} (\bibinfo {year} {2012})}\BibitemShut {NoStop}%
\bibitem [{\citenamefont {Brown}\ and\ \citenamefont {Forsyth}(1995)}]{Mn5Si3-magnetic_state3}%
  \BibitemOpen
  \bibfield  {author} {\bibinfo {author} {\bibfnamefont {P.}~\bibnamefont {Brown}}\ and\ \bibinfo {author} {\bibfnamefont {J.}~\bibnamefont {Forsyth}},\ }\bibfield  {title} {\bibinfo {title} {Antiferromagnetism in Mn$_5$Si$_3$: the magnetic structure of the af2 phase at 70 k},\ }\href@noop {} {\bibfield  {journal} {\bibinfo  {journal} {Journal of Physics: Condensed Matter}\ }\textbf {\bibinfo {volume} {7}},\ \bibinfo {pages} {7619} (\bibinfo {year} {1995})}\BibitemShut {NoStop}%
\bibitem [{\citenamefont {Biniskos}\ \emph {et~al.}(2022)\citenamefont {Biniskos}, \citenamefont {dos Santos}, \citenamefont {Schmalzl}, \citenamefont {Raymond}, \citenamefont {dos Santos~Dias}, \citenamefont {Persson}, \citenamefont {Marzari}, \citenamefont {Bl\"ugel}, \citenamefont {Lounis},\ and\ \citenamefont {Br\"uckel}}]{Mn5Si3-magnetic_state4}%
  \BibitemOpen
  \bibfield  {author} {\bibinfo {author} {\bibfnamefont {N.}~\bibnamefont {Biniskos}}, \bibinfo {author} {\bibfnamefont {F.~J.}\ \bibnamefont {dos Santos}}, \bibinfo {author} {\bibfnamefont {K.}~\bibnamefont {Schmalzl}}, \bibinfo {author} {\bibfnamefont {S.}~\bibnamefont {Raymond}}, \bibinfo {author} {\bibfnamefont {M.}~\bibnamefont {dos Santos~Dias}}, \bibinfo {author} {\bibfnamefont {J.}~\bibnamefont {Persson}}, \bibinfo {author} {\bibfnamefont {N.}~\bibnamefont {Marzari}}, \bibinfo {author} {\bibfnamefont {S.}~\bibnamefont {Bl\"ugel}}, \bibinfo {author} {\bibfnamefont {S.}~\bibnamefont {Lounis}},\ and\ \bibinfo {author} {\bibfnamefont {T.}~\bibnamefont {Br\"uckel}},\ }\bibfield  {title} {\bibinfo {title} {Complex magnetic structure and spin waves of the noncollinear antiferromagnet ${\mathrm{Mn}}_{5}{\mathrm{Si}}_{3}$},\ }\href {https://doi.org/10.1103/PhysRevB.105.104404} {\bibfield  {journal} {\bibinfo  {journal} {Phys. Rev. B}\ }\textbf {\bibinfo {volume} {105}},\ \bibinfo {pages} {104404} (\bibinfo
  {year} {2022})}\BibitemShut {NoStop}%
\bibitem [{\citenamefont {Urata}\ \emph {et~al.}(2024)\citenamefont {Urata}, \citenamefont {Hattori},\ and\ \citenamefont {Ikuta}}]{CrSb-multicarrier}%
  \BibitemOpen
  \bibfield  {author} {\bibinfo {author} {\bibfnamefont {T.}~\bibnamefont {Urata}}, \bibinfo {author} {\bibfnamefont {W.}~\bibnamefont {Hattori}},\ and\ \bibinfo {author} {\bibfnamefont {H.}~\bibnamefont {Ikuta}},\ }\bibfield  {title} {\bibinfo {title} {High mobility charge transport in a multicarrier altermagnet CrSb},\ }\href@noop {} {\bibfield  {journal} {\bibinfo  {journal} {Physical Review Materials}\ }\textbf {\bibinfo {volume} {8}},\ \bibinfo {pages} {084412} (\bibinfo {year} {2024})}\BibitemShut {NoStop}%
\bibitem [{\citenamefont {Lu}\ \emph {et~al.}(2024)\citenamefont {Lu}, \citenamefont {Feng}, \citenamefont {Wang}, \citenamefont {Chen}, \citenamefont {Lin}, \citenamefont {Liang}, \citenamefont {Liu}, \citenamefont {Feng}, \citenamefont {Yamagami}, \citenamefont {Liu} \emph {et~al.}}]{CrSb-fermi-arc}%
  \BibitemOpen
  \bibfield  {author} {\bibinfo {author} {\bibfnamefont {W.}~\bibnamefont {Lu}}, \bibinfo {author} {\bibfnamefont {S.}~\bibnamefont {Feng}}, \bibinfo {author} {\bibfnamefont {Y.}~\bibnamefont {Wang}}, \bibinfo {author} {\bibfnamefont {D.}~\bibnamefont {Chen}}, \bibinfo {author} {\bibfnamefont {Z.}~\bibnamefont {Lin}}, \bibinfo {author} {\bibfnamefont {X.}~\bibnamefont {Liang}}, \bibinfo {author} {\bibfnamefont {S.}~\bibnamefont {Liu}}, \bibinfo {author} {\bibfnamefont {W.}~\bibnamefont {Feng}}, \bibinfo {author} {\bibfnamefont {K.}~\bibnamefont {Yamagami}}, \bibinfo {author} {\bibfnamefont {J.}~\bibnamefont {Liu}}, \emph {et~al.},\ }\bibfield  {title} {\bibinfo {title} {Observation of surface fermi arcs in altermagnetic weyl semimetal CrSb},\ }\href@noop {} {\bibfield  {journal} {\bibinfo  {journal} {arXiv preprint arXiv:2407.13497}\ } (\bibinfo {year} {2024})}\BibitemShut {NoStop}%
\bibitem [{\citenamefont {Zeng}\ \emph {et~al.}(2024)\citenamefont {Zeng}, \citenamefont {Zhu}, \citenamefont {Zhu}, \citenamefont {Liu}, \citenamefont {Ma}, \citenamefont {Hao}, \citenamefont {Liu}, \citenamefont {Qu}, \citenamefont {Yang}, \citenamefont {Jiang} \emph {et~al.}}]{CrSb-ARPES1}%
  \BibitemOpen
  \bibfield  {author} {\bibinfo {author} {\bibfnamefont {M.}~\bibnamefont {Zeng}}, \bibinfo {author} {\bibfnamefont {M.-Y.}\ \bibnamefont {Zhu}}, \bibinfo {author} {\bibfnamefont {Y.-P.}\ \bibnamefont {Zhu}}, \bibinfo {author} {\bibfnamefont {X.-R.}\ \bibnamefont {Liu}}, \bibinfo {author} {\bibfnamefont {X.-M.}\ \bibnamefont {Ma}}, \bibinfo {author} {\bibfnamefont {Y.-J.}\ \bibnamefont {Hao}}, \bibinfo {author} {\bibfnamefont {P.}~\bibnamefont {Liu}}, \bibinfo {author} {\bibfnamefont {G.}~\bibnamefont {Qu}}, \bibinfo {author} {\bibfnamefont {Y.}~\bibnamefont {Yang}}, \bibinfo {author} {\bibfnamefont {Z.}~\bibnamefont {Jiang}}, \emph {et~al.},\ }\bibfield  {title} {\bibinfo {title} {Observation of spin splitting in room-temperature metallic antiferromagnet CrSb},\ }\href@noop {} {\bibfield  {journal} {\bibinfo  {journal} {Advanced Science}\ ,\ \bibinfo {pages} {2406529}} (\bibinfo {year} {2024})}\BibitemShut {NoStop}%
\bibitem [{\citenamefont {Ding}\ \emph {et~al.}(2024)\citenamefont {Ding}, \citenamefont {Jiang}, \citenamefont {Chen}, \citenamefont {Tao}, \citenamefont {Liu}, \citenamefont {Liu}, \citenamefont {Li}, \citenamefont {Liu}, \citenamefont {Yang}, \citenamefont {Zhang} \emph {et~al.}}]{CrSb-ARPES2}%
  \BibitemOpen
  \bibfield  {author} {\bibinfo {author} {\bibfnamefont {J.}~\bibnamefont {Ding}}, \bibinfo {author} {\bibfnamefont {Z.}~\bibnamefont {Jiang}}, \bibinfo {author} {\bibfnamefont {X.}~\bibnamefont {Chen}}, \bibinfo {author} {\bibfnamefont {Z.}~\bibnamefont {Tao}}, \bibinfo {author} {\bibfnamefont {Z.}~\bibnamefont {Liu}}, \bibinfo {author} {\bibfnamefont {J.}~\bibnamefont {Liu}}, \bibinfo {author} {\bibfnamefont {T.}~\bibnamefont {Li}}, \bibinfo {author} {\bibfnamefont {J.}~\bibnamefont {Liu}}, \bibinfo {author} {\bibfnamefont {Y.}~\bibnamefont {Yang}}, \bibinfo {author} {\bibfnamefont {R.}~\bibnamefont {Zhang}}, \emph {et~al.},\ }\bibfield  {title} {\bibinfo {title} {Large band-splitting in $ g $-wave type altermagnet CrSb},\ }\href@noop {} {\bibfield  {journal} {\bibinfo  {journal} {arXiv preprint arXiv:2405.12687}\ } (\bibinfo {year} {2024})}\BibitemShut {NoStop}%
\bibitem [{\citenamefont {Yang}\ \emph {et~al.}(2024)\citenamefont {Yang}, \citenamefont {Li}, \citenamefont {Yang}, \citenamefont {Li}, \citenamefont {Zheng}, \citenamefont {Zhu}, \citenamefont {Cao}, \citenamefont {Zhao}, \citenamefont {Zhang}, \citenamefont {Ye} \emph {et~al.}}]{CrSb-ARPES3}%
  \BibitemOpen
  \bibfield  {author} {\bibinfo {author} {\bibfnamefont {G.}~\bibnamefont {Yang}}, \bibinfo {author} {\bibfnamefont {Z.}~\bibnamefont {Li}}, \bibinfo {author} {\bibfnamefont {S.}~\bibnamefont {Yang}}, \bibinfo {author} {\bibfnamefont {J.}~\bibnamefont {Li}}, \bibinfo {author} {\bibfnamefont {H.}~\bibnamefont {Zheng}}, \bibinfo {author} {\bibfnamefont {W.}~\bibnamefont {Zhu}}, \bibinfo {author} {\bibfnamefont {S.}~\bibnamefont {Cao}}, \bibinfo {author} {\bibfnamefont {W.}~\bibnamefont {Zhao}}, \bibinfo {author} {\bibfnamefont {J.}~\bibnamefont {Zhang}}, \bibinfo {author} {\bibfnamefont {M.}~\bibnamefont {Ye}}, \emph {et~al.},\ }\bibfield  {title} {\bibinfo {title} {Three-dimensional mapping and electronic origin of large altermagnetic splitting near fermi level in CrSb},\ }\href@noop {} {\bibfield  {journal} {\bibinfo  {journal} {arXiv preprint arXiv:2405.12575}\ } (\bibinfo {year} {2024})}\BibitemShut {NoStop}%
\bibitem [{\citenamefont {Reimers}\ \emph {et~al.}(2024)\citenamefont {Reimers}, \citenamefont {Odenbreit}, \citenamefont {{\v{S}}mejkal}, \citenamefont {Strocov}, \citenamefont {Constantinou}, \citenamefont {Hellenes}, \citenamefont {Jaeschke~Ubiergo}, \citenamefont {Campos}, \citenamefont {Bharadwaj}, \citenamefont {Chakraborty} \emph {et~al.}}]{CrSb-ARPES4}%
  \BibitemOpen
  \bibfield  {author} {\bibinfo {author} {\bibfnamefont {S.}~\bibnamefont {Reimers}}, \bibinfo {author} {\bibfnamefont {L.}~\bibnamefont {Odenbreit}}, \bibinfo {author} {\bibfnamefont {L.}~\bibnamefont {{\v{S}}mejkal}}, \bibinfo {author} {\bibfnamefont {V.~N.}\ \bibnamefont {Strocov}}, \bibinfo {author} {\bibfnamefont {P.}~\bibnamefont {Constantinou}}, \bibinfo {author} {\bibfnamefont {A.~B.}\ \bibnamefont {Hellenes}}, \bibinfo {author} {\bibfnamefont {R.}~\bibnamefont {Jaeschke~Ubiergo}}, \bibinfo {author} {\bibfnamefont {W.~H.}\ \bibnamefont {Campos}}, \bibinfo {author} {\bibfnamefont {V.~K.}\ \bibnamefont {Bharadwaj}}, \bibinfo {author} {\bibfnamefont {A.}~\bibnamefont {Chakraborty}}, \emph {et~al.},\ }\bibfield  {title} {\bibinfo {title} {Direct observation of altermagnetic band splitting in CrSb thin films},\ }\href@noop {} {\bibfield  {journal} {\bibinfo  {journal} {Nature Communications}\ }\textbf {\bibinfo {volume} {15}},\ \bibinfo {pages} {2116} (\bibinfo {year} {2024})}\BibitemShut {NoStop}%
\bibitem [{\citenamefont {Takei}\ \emph {et~al.}(1963)\citenamefont {Takei}, \citenamefont {Cox},\ and\ \citenamefont {Shirane}}]{CrSb-Neel-temperature}%
  \BibitemOpen
  \bibfield  {author} {\bibinfo {author} {\bibfnamefont {W.}~\bibnamefont {Takei}}, \bibinfo {author} {\bibfnamefont {D.~E.}\ \bibnamefont {Cox}},\ and\ \bibinfo {author} {\bibfnamefont {G.}~\bibnamefont {Shirane}},\ }\bibfield  {title} {\bibinfo {title} {Magnetic structures in the MnSb-CrSb system},\ }\href@noop {} {\bibfield  {journal} {\bibinfo  {journal} {Physical Review}\ }\textbf {\bibinfo {volume} {129}},\ \bibinfo {pages} {2008} (\bibinfo {year} {1963})}\BibitemShut {NoStop}%
\bibitem [{\citenamefont {Seemann}\ \emph {et~al.}(2015)\citenamefont {Seemann}, \citenamefont {K{\"o}dderitzsch}, \citenamefont {Wimmer},\ and\ \citenamefont {Ebert}}]{2015symmetry}%
  \BibitemOpen
  \bibfield  {author} {\bibinfo {author} {\bibfnamefont {M.}~\bibnamefont {Seemann}}, \bibinfo {author} {\bibfnamefont {D.}~\bibnamefont {K{\"o}dderitzsch}}, \bibinfo {author} {\bibfnamefont {S.}~\bibnamefont {Wimmer}},\ and\ \bibinfo {author} {\bibfnamefont {H.}~\bibnamefont {Ebert}},\ }\bibfield  {title} {\bibinfo {title} {Symmetry-imposed shape of linear response tensors},\ }\href@noop {} {\bibfield  {journal} {\bibinfo  {journal} {Physical Review B}\ }\textbf {\bibinfo {volume} {92}},\ \bibinfo {pages} {155138} (\bibinfo {year} {2015})}\BibitemShut {NoStop}%
\bibitem [{web()}]{website}%
  \BibitemOpen
  \href@noop {} {}\bibinfo {howpublished} {\url{https://bitbucket.org/zeleznyj/linear-response-symmetry/src/master/}}\BibitemShut {NoStop}%
\bibitem [{\citenamefont {Kresse}\ and\ \citenamefont {Furthm{\"u}ller}(1996)}]{vasp}%
  \BibitemOpen
  \bibfield  {author} {\bibinfo {author} {\bibfnamefont {G.}~\bibnamefont {Kresse}}\ and\ \bibinfo {author} {\bibfnamefont {J.}~\bibnamefont {Furthm{\"u}ller}},\ }\bibfield  {title} {\bibinfo {title} {Efficient iterative schemes for ab initio total-energy calculations using a plane-wave basis set},\ }\href@noop {} {\bibfield  {journal} {\bibinfo  {journal} {Physical review B}\ }\textbf {\bibinfo {volume} {54}},\ \bibinfo {pages} {11169} (\bibinfo {year} {1996})}\BibitemShut {NoStop}%
\bibitem [{\citenamefont {Perdew}\ \emph {et~al.}(1996)\citenamefont {Perdew}, \citenamefont {Burke},\ and\ \citenamefont {Ernzerhof}}]{pbe}%
  \BibitemOpen
  \bibfield  {author} {\bibinfo {author} {\bibfnamefont {J.~P.}\ \bibnamefont {Perdew}}, \bibinfo {author} {\bibfnamefont {K.}~\bibnamefont {Burke}},\ and\ \bibinfo {author} {\bibfnamefont {M.}~\bibnamefont {Ernzerhof}},\ }\bibfield  {title} {\bibinfo {title} {Generalized gradient approximation made simple},\ }\href@noop {} {\bibfield  {journal} {\bibinfo  {journal} {Physical review letters}\ }\textbf {\bibinfo {volume} {77}},\ \bibinfo {pages} {3865} (\bibinfo {year} {1996})}\BibitemShut {NoStop}%
\bibitem [{\citenamefont {Mostofi}\ \emph {et~al.}(2008)\citenamefont {Mostofi}, \citenamefont {Yates}, \citenamefont {Lee}, \citenamefont {Souza}, \citenamefont {Vanderbilt},\ and\ \citenamefont {Marzari}}]{wannier90}%
  \BibitemOpen
  \bibfield  {author} {\bibinfo {author} {\bibfnamefont {A.~A.}\ \bibnamefont {Mostofi}}, \bibinfo {author} {\bibfnamefont {J.~R.}\ \bibnamefont {Yates}}, \bibinfo {author} {\bibfnamefont {Y.-S.}\ \bibnamefont {Lee}}, \bibinfo {author} {\bibfnamefont {I.}~\bibnamefont {Souza}}, \bibinfo {author} {\bibfnamefont {D.}~\bibnamefont {Vanderbilt}},\ and\ \bibinfo {author} {\bibfnamefont {N.}~\bibnamefont {Marzari}},\ }\bibfield  {title} {\bibinfo {title} {wannier90: A tool for obtaining maximally-localised wannier functions},\ }\href@noop {} {\bibfield  {journal} {\bibinfo  {journal} {Computer physics communications}\ }\textbf {\bibinfo {volume} {178}},\ \bibinfo {pages} {685} (\bibinfo {year} {2008})}\BibitemShut {NoStop}%
\bibitem [{\citenamefont {Noky}\ \emph {et~al.}(2020)\citenamefont {Noky}, \citenamefont {Zhang}, \citenamefont {Gooth}, \citenamefont {Felser},\ and\ \citenamefont {Sun}}]{ANE-npj-database}%
  \BibitemOpen
  \bibfield  {author} {\bibinfo {author} {\bibfnamefont {J.}~\bibnamefont {Noky}}, \bibinfo {author} {\bibfnamefont {Y.}~\bibnamefont {Zhang}}, \bibinfo {author} {\bibfnamefont {J.}~\bibnamefont {Gooth}}, \bibinfo {author} {\bibfnamefont {C.}~\bibnamefont {Felser}},\ and\ \bibinfo {author} {\bibfnamefont {Y.}~\bibnamefont {Sun}},\ }\bibfield  {title} {\bibinfo {title} {Giant anomalous Hall and Nernst effect in magnetic cubic heusler compounds},\ }\href@noop {} {\bibfield  {journal} {\bibinfo  {journal} {npj Computational Materials}\ }\textbf {\bibinfo {volume} {6}},\ \bibinfo {pages} {77} (\bibinfo {year} {2020})}\BibitemShut {NoStop}%
\bibitem [{\citenamefont {Park}\ \emph {et~al.}(2011)\citenamefont {Park}, \citenamefont {Wunderlich}, \citenamefont {Mart{\'\i}}, \citenamefont {Hol{\`y}}, \citenamefont {Kurosaki}, \citenamefont {Yamada}, \citenamefont {Yamamoto}, \citenamefont {Nishide}, \citenamefont {Hayakawa}, \citenamefont {Takahashi} \emph {et~al.}}]{park2011}%
  \BibitemOpen
  \bibfield  {author} {\bibinfo {author} {\bibfnamefont {B.~G.}\ \bibnamefont {Park}}, \bibinfo {author} {\bibfnamefont {J.}~\bibnamefont {Wunderlich}}, \bibinfo {author} {\bibfnamefont {X.}~\bibnamefont {Mart{\'\i}}}, \bibinfo {author} {\bibfnamefont {V.}~\bibnamefont {Hol{\`y}}}, \bibinfo {author} {\bibfnamefont {Y.}~\bibnamefont {Kurosaki}}, \bibinfo {author} {\bibfnamefont {M.}~\bibnamefont {Yamada}}, \bibinfo {author} {\bibfnamefont {H.}~\bibnamefont {Yamamoto}}, \bibinfo {author} {\bibfnamefont {A.}~\bibnamefont {Nishide}}, \bibinfo {author} {\bibfnamefont {J.}~\bibnamefont {Hayakawa}}, \bibinfo {author} {\bibfnamefont {H.}~\bibnamefont {Takahashi}}, \emph {et~al.},\ }\bibfield  {title} {\bibinfo {title} {A spin-valve-like magnetoresistance of an antiferromagnet-based tunnel junction},\ }\href@noop {} {\bibfield  {journal} {\bibinfo  {journal} {Nature materials}\ }\textbf {\bibinfo {volume} {10}},\ \bibinfo {pages} {347} (\bibinfo {year} {2011})}\BibitemShut {NoStop}%
\bibitem [{\citenamefont {Zhou}\ \emph {et~al.}(2025)\citenamefont {Zhou}, \citenamefont {Cheng}, \citenamefont {Hu}, \citenamefont {Chu}, \citenamefont {Bai}, \citenamefont {Han}, \citenamefont {Liu}, \citenamefont {Pan},\ and\ \citenamefont {Song}}]{CrSb_nature}%
  \BibitemOpen
  \bibfield  {author} {\bibinfo {author} {\bibfnamefont {Z.}~\bibnamefont {Zhou}}, \bibinfo {author} {\bibfnamefont {X.}~\bibnamefont {Cheng}}, \bibinfo {author} {\bibfnamefont {M.}~\bibnamefont {Hu}}, \bibinfo {author} {\bibfnamefont {R.}~\bibnamefont {Chu}}, \bibinfo {author} {\bibfnamefont {H.}~\bibnamefont {Bai}}, \bibinfo {author} {\bibfnamefont {L.}~\bibnamefont {Han}}, \bibinfo {author} {\bibfnamefont {J.}~\bibnamefont {Liu}}, \bibinfo {author} {\bibfnamefont {F.}~\bibnamefont {Pan}},\ and\ \bibinfo {author} {\bibfnamefont {C.}~\bibnamefont {Song}},\ }\bibfield  {title} {\bibinfo {title} {Manipulation of the altermagnetic order in crsb via crystal symmetry},\ }\href@noop {} {\bibfield  {journal} {\bibinfo  {journal} {Nature}\ ,\ \bibinfo {pages} {1}} (\bibinfo {year} {2025})}\BibitemShut {NoStop}%
\end{thebibliography}
%


\begin{figure}
\includegraphics[width=3.5 in]{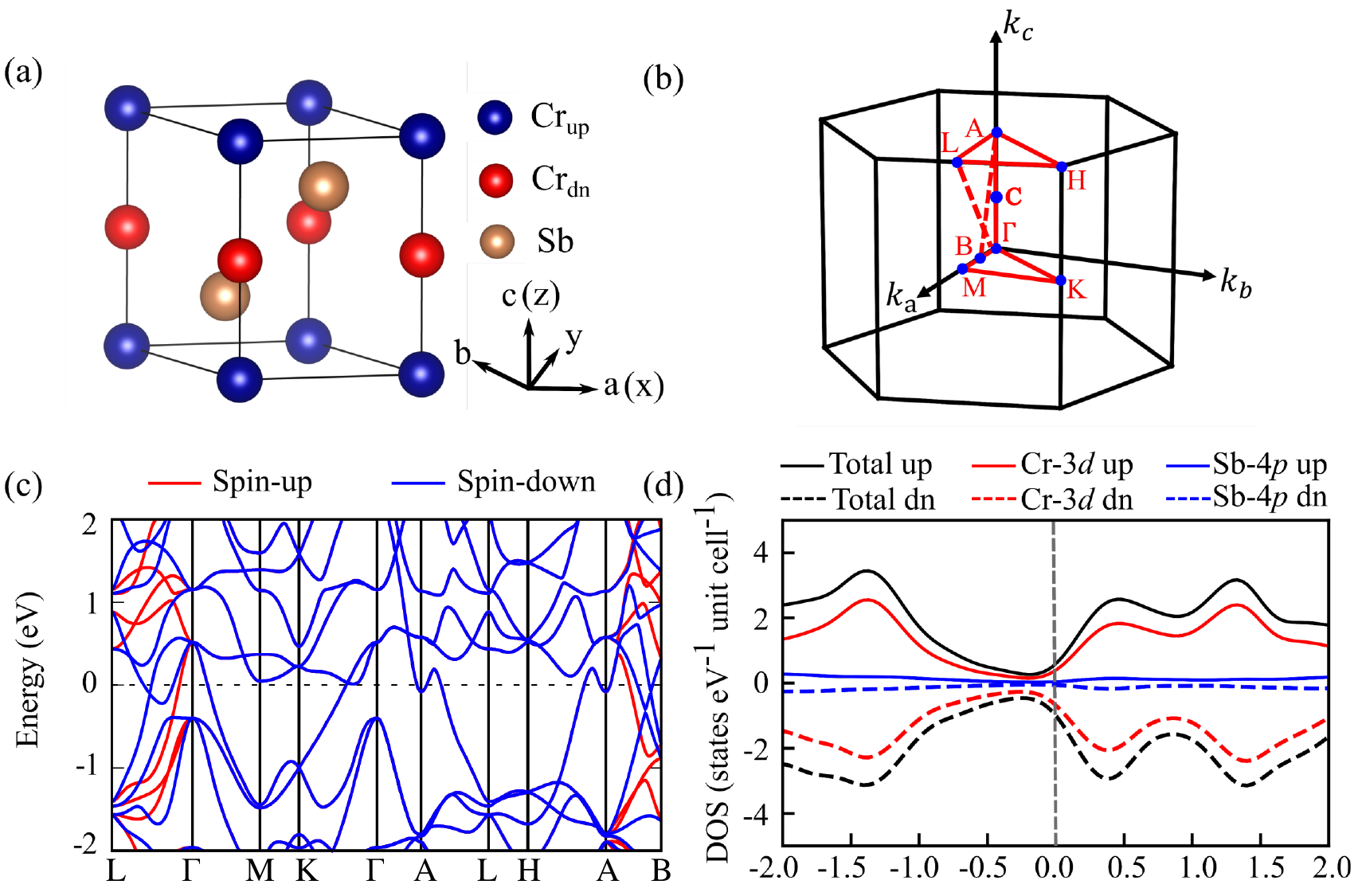}
\caption{\label{fig:wide}Crystal structure and electronic structure of CrSb. (a) Crystal structure of CrSb with 
space group $P6_3/mmc$ (No. 194). The blue and red spheres 
represent Cr atoms with opposite magnetic moment, and the brown spheres represent Sb atoms. (b) Brillouin zone of the CrSb 
structure. 
(c) Band structure of CrSb, with red and blue lines 
denoting spin-up and spin-down bands, respectively. 
(d) The total electronic density of states and partial orbit-resolved electronic density of states with spin up and spin down character in the altermagnetic state.}  
\end{figure}

\begin{figure}
\includegraphics[width=3.5in]{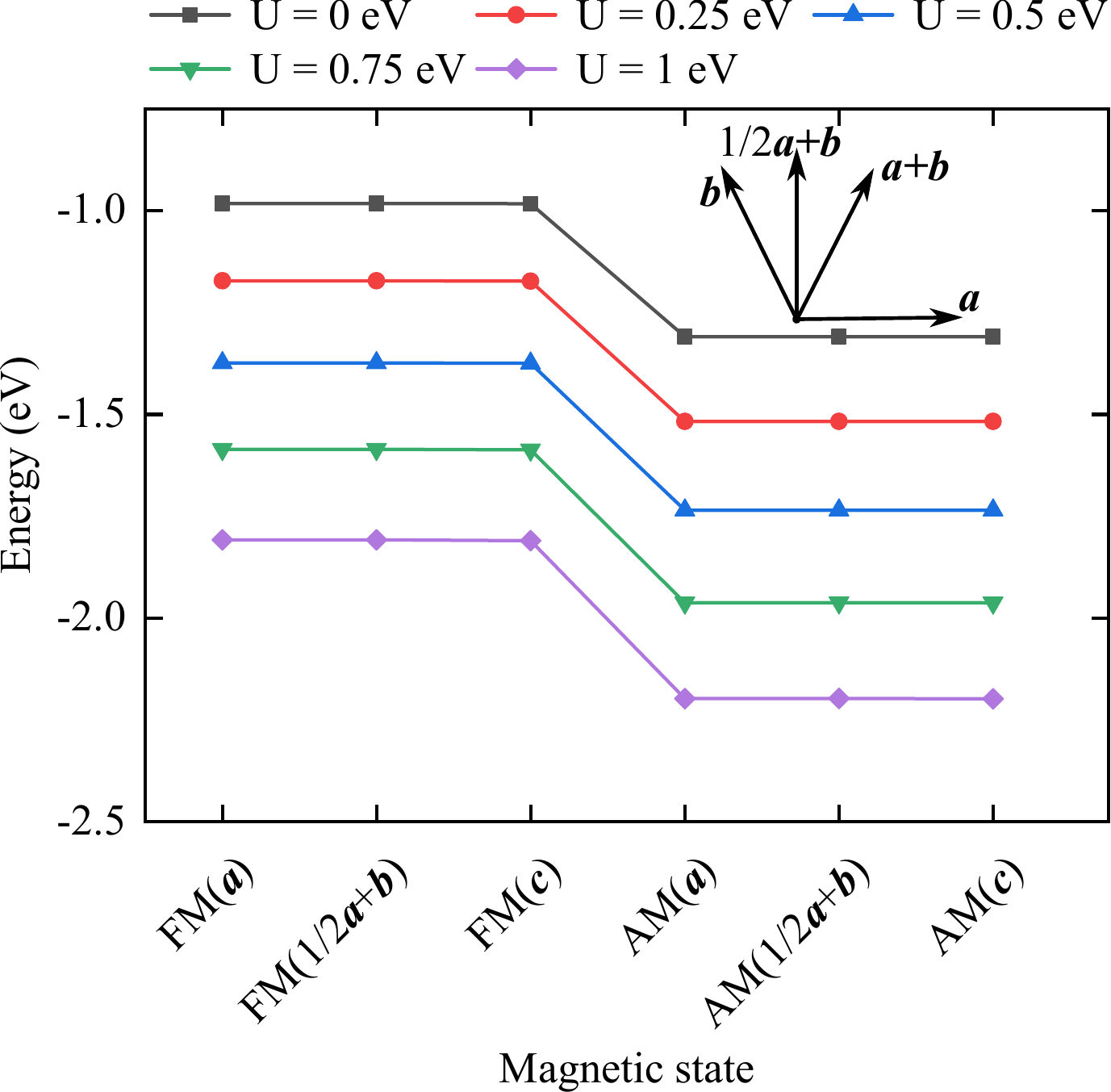}
\caption{\label{fig:wide}Dependence of total energy on the 
orientation of the magnetic moment in the ferromagnetic 
state and the N\'eel vector in the altermagnetic state. 
The total energy of the nonmagnetic state is aligned to zero. 
Different lines represent results obtained with varying 
Coulomb repulsion energies, ranging from 0 to 1 eV. The inset depicts the vector orientations, with \textbf{\textit{c}} indicating the out-of-plane direction.}
\end{figure} 

\begin{figure}[bhtp] 
\centering
\includegraphics[width=3.5in]{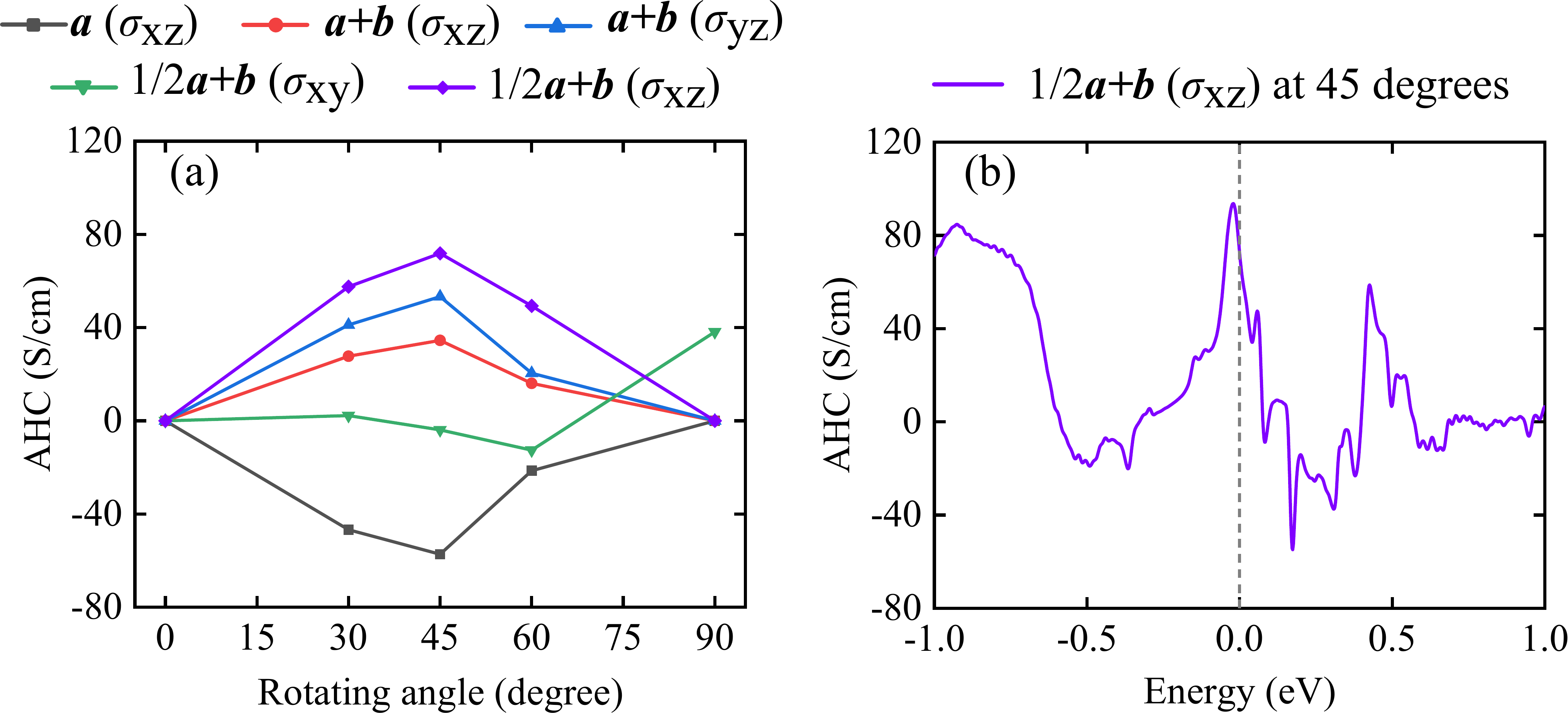}
\caption{Rotating angle-dependent and energy-dependent AHC. (a) Dependence of AHC on the N\'eel vector as it rotates from $\textbf{\textit{c}}$ 
to the orientations indicated in the legend. The letters 
in parentheses denote the nonzero elements of the AHC 
tensor $\sigma$ corresponding to N\'eel vector. Note that there are two independent 
nonzero AHC tensor elements when the N\'eel vector rotates 
from \textbf{\textit{c}} to \textbf{\textit{a}}+\textbf{\textit{b}} and 
$\frac{1}{2}$\textbf{\textit{a}}+\textbf{\textit{b}}. (b) Energy dependence 
of $\sigma_{xz}$ with the N\'eel vector aligned at 45 
degrees from \textbf{\textit{c}}, as it rotates from 
\textbf{\textit{c}} to $\frac{1}{2}$\textbf{\textit{a}}+\textbf{\textit{b}}.} 
\end{figure}

\begin{figure}[htbp]
\includegraphics[width=3.5in]{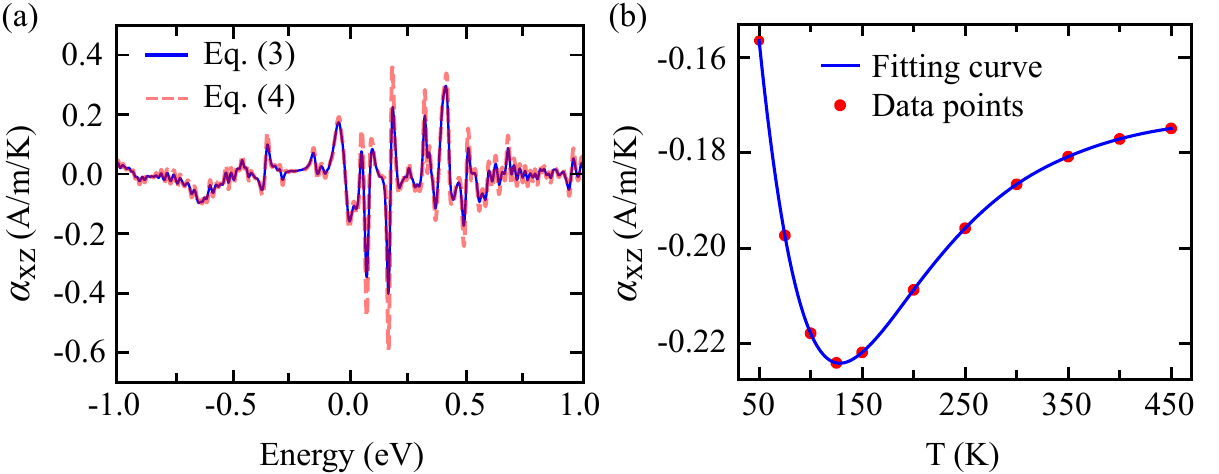}
\caption{\label{fig:4}Energy-dependent and temperature-dependent ANC. (a) ANC $\alpha_{xz}$ in CrSb at 50 K, with the Néel vector 
aligned at 45 degrees from \textbf{\textit{c}} during rotation 
from \textbf{\textit{c}} to $\frac{1}{2}$\textbf{\textit{a}}+\textbf{\textit{b}}. 
Results are calculated using both the BC formalism in Eq. (3) (blue solid line) and the 
Mott relation in Eq. (4) (pink dashed line).
(b) Temperature dependence of ANC $\alpha_{xz}$, with red points 
representing the original calculation results and the blue 
line indicating the fitting curve.
}
\end{figure} 

\begin{figure}[pbht]
\includegraphics[width=3.5 in]{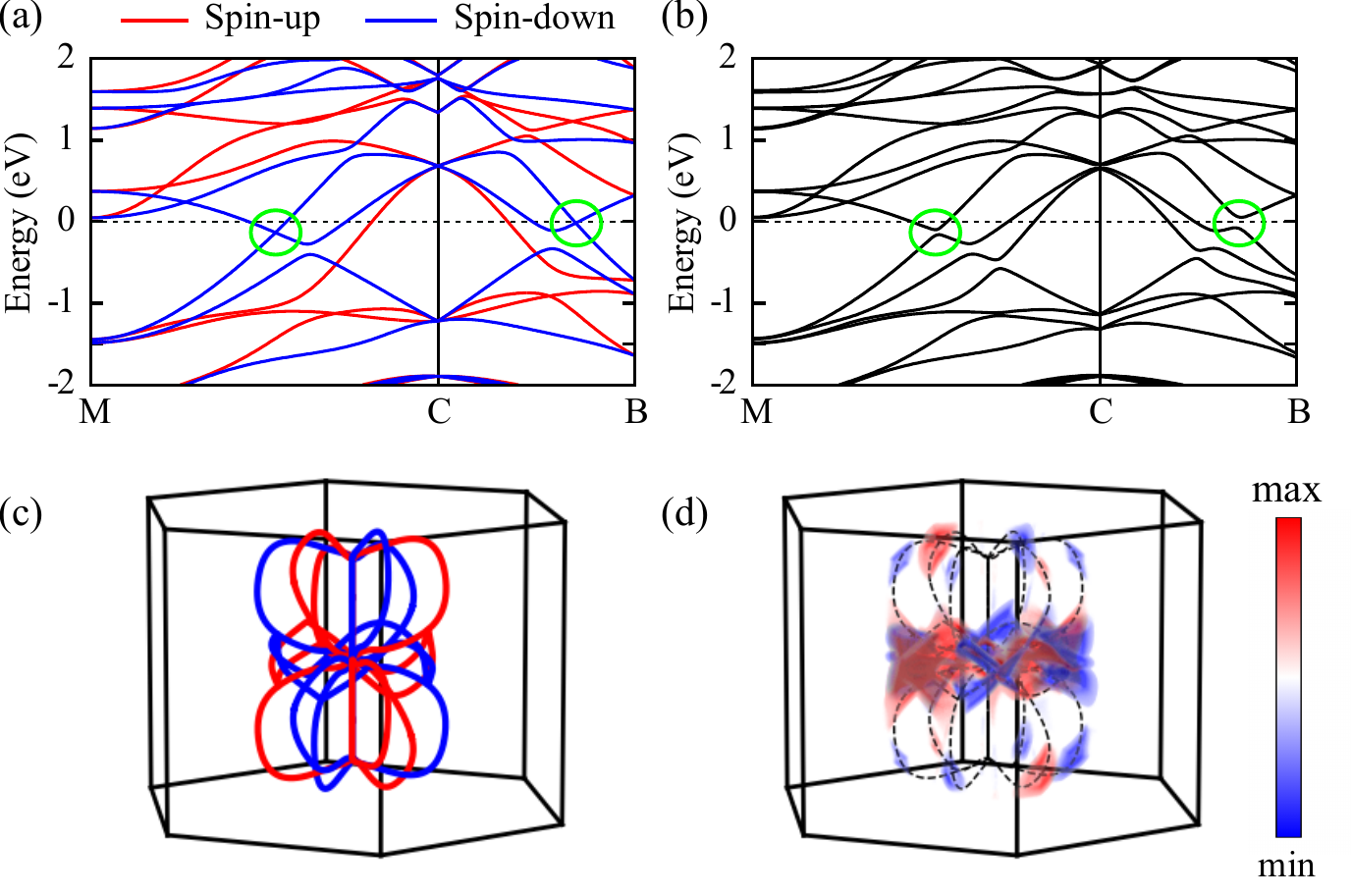}
\caption{\label{fig:wide}Berry curvature distribution in CrSb. Band structures with the N\'eel vector aligned at 45 degrees from $\textbf{\textit{c}}$ during its rotation from $\textbf{\textit{c}}$ to $\frac{1}{2}$\textbf{\textit{a}}+\textbf{\textit{b}}, calculated 
(a) without and (b) with SOC. The green circles in (a) 
indicate the crossing points formed by the 25th and 26th 
spin-down bands, red and blue represent spin-up and 
spin-down bands, respectively. (c) Nodal rings formed 
by the 25th and 26th bands, showing both spin-up (red) 
and spin-down (blue) characters. (d) Distribution of 
Berry curvature and nodal rings in the Brillouin zone, 
with nodal rings represented by black dashed lines.}
\end{figure}

\begin{figure}[pbht]
\includegraphics[width=3.5 in]{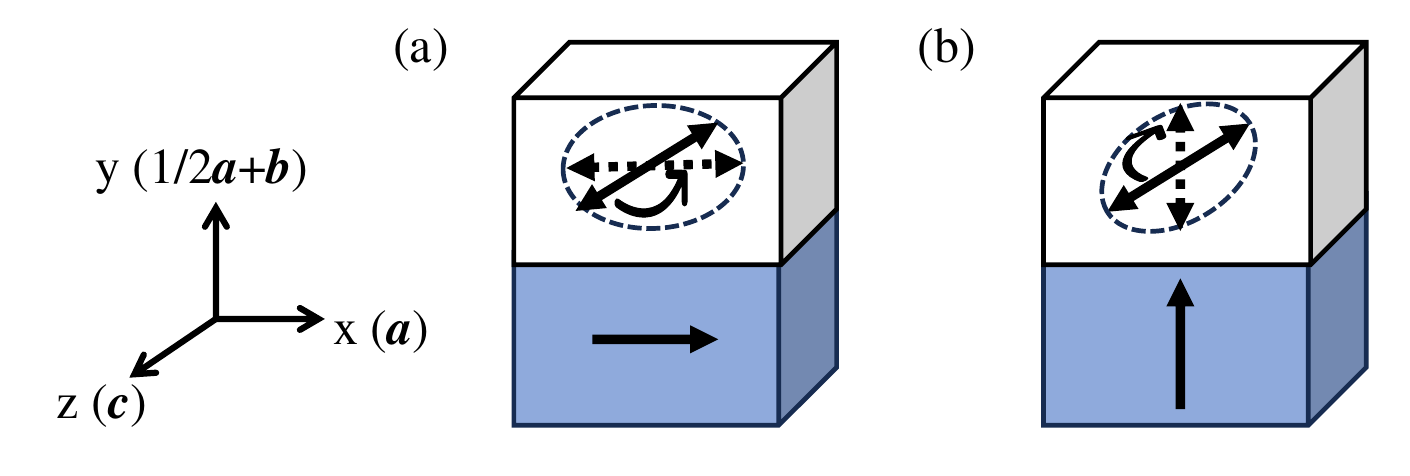}
\caption{\label{fig:wide} {Experimental setups where CrSb films are grown on the ferromagnetic substrates for the detection of $\sigma_{xz}$ as N\'eel vectors rotate from $\textbf{\textit{c}}$ to (a) $\textbf{\textit{a}}$ and (b) $\frac{1}{2}\textbf{\textit{a}} + \textbf{\textit{b}}$. The arrows in the substrates and CrSb films indicate the orientations of the ferromagnetic moments and N\'eel vectors, respectively. The substrate in (a) can be a conventional ferromagnet, which meets the common expression $j^{\text{AHE}} \sim M \times E$, the substrate in (b) can be an insulating ferromagnet, such as CrI$_3$ and Cr$_2$Ge$_2$Te$_6$.}}
\end{figure}

\begin{table*}[!htbp]
\caption{\label{tab:table1}%
Total energy of CrSb in units of meV, calculated for 
different N\'eel vectors and on-site Coulomb repulsion 
energies $U$. The total energy with the N\'eel vector 
aligned along \textbf{\textit{c}} is aligned to zero.}
\begin{ruledtabular}
\begin{tabular}{ccccccc}   
   & $U$ = 0 eV	& $U$ = 0.25 eV	& $U$ = 0.5 eV	& $U$= 0.75 eV	& $U$ = 1 eV  \\
  \hline
  \textbf{\textit{a}} & 0.008 &0.185 &0.064 &-0.033 &0.335  \\
  $\frac{1}{2}$\textbf{\textit{a}}+\textbf{\textit{b}} & 0.164 &0.153 &0.230 &0.119 &0.567  \\
  \textbf{\textit{c}} & 0 & 0 & 0 & 0 & 0 
\end{tabular}
\end{ruledtabular}
\end{table*} 

\begin{table}[!htbp]
\caption{\label{tab:table2}%
ANC $\alpha$ at 300 K, in units of A/m/K, as the N\'eel vector rotates from 
\textbf{\textit{c}} to the orientations listed in the 
first column. Note that there are two independent 
nonzero ANC tensor elements when the N\'eel vector 
rotates from \textbf{\textit{c}} to \textbf{\textit{a}}+\textbf{\textit{b}} 
and $\frac{1}{2}$\textbf{\textit{a}}+\textbf{\textit{b}}. The rotation 
angles range from 0 to 90 degrees.}
\begin{ruledtabular}
\begin{tabular}{ccccccc}
   &  & 0	&30	&45	&60	&90 \\
  \hline
  \textbf{\textit{a}} & $\alpha_{xz}$ & 0 &0.07 &0.10 &0.05 &0   \\
  \textbf{\textit{a}}+\textbf{\textit{b}} & $\alpha_{xz}$ & 0 & -0.02 &-0.02 &0.02  &0  \\
  \textbf{\textit{a}}+\textbf{\textit{b}} & $\alpha_{yz}$ & 0 & -0.06 &-0.07 &-0.02  &0  \\
  $\frac{1}{2}$\textbf{\textit{a}}+\textbf{\textit{b}} & $\alpha_{xy}$ & 0 & -0.06 &-0.07 &-0.08  &0.06  \\
  $\frac{1}{2}$\textbf{\textit{a}}+\textbf{\textit{b}} & $\alpha_{xz}$ & 0 & -0.11 &-0.19 &-0.18  &0  
\end{tabular}
\end{ruledtabular}
\end{table} 

\end{document}